\documentclass[aps,nofootinbib,floatfix,showpacs,preprintnumbers,prd]{revtex4}
\usepackage{graphicx, epsfig, bm, amsmath}
\usepackage{color}
\usepackage{float}
\usepackage{hyperref}

\usepackage{wasysym}

\begin{document}
\preprint{IRFU-15-01}


\definecolor{rosso}{cmyk}{0,1,1,0.4}
\newcommand{\xxx}[1]{{\bf\color{rosso} #1}\marginpar{$\bullet$}}

\newcommand{\beq}{\begin{equation}}
\newcommand{\eeq}{\end{equation}}

\newcommand{\mDM}{m_\text{DM}}
\newcommand{\sigmav}{\langle \sigma v \rangle}
\newcommand{\ud}{\text{d}}
\newcommand{\DM}{dark matter}

\def\LHC{{\sf LHC}}

\def\FERMI{{\sf Fermi-LAT}}
\def\HESS{{\sf H.E.S.S.}}
\def\MAGIC{{\sf MAGIC}}
\def\VERITAS{{\sf VERITAS}}
\def\CTA{{\sf CTA}}

\title{Prospects for Annihilating Dark Matter in the inner Galactic halo \\ by the Cherenkov Telescope Array}
\author{Valentin Lefranc}
\email{valentin.lefranc@cea.fr}
\affiliation{CEA Irfu, Service de Physique des Particules, Centre de Saclay,
Gif-sur-Yvette 91190, France} 

\author{Emmanuel Moulin}
\email{emmanuel.moulin@cea.fr}
\affiliation{CEA Irfu, Service de Physique des Particules, Centre de Saclay,
Gif-sur-Yvette 91190, France} 

\author{Paolo Panci}
\email{panci@iap.fr}
\affiliation{Institut d'Astrophysique de Paris, UMR 7095 CNRS, Universit\'e Pierre et Marie Curie, 98 bis Boulevard Arago, Paris 75014, France}

\author{Joseph Silk}
\email{silk@iap.fr}
\affiliation{Institut d'Astrophysique de Paris, UMR 7095 CNRS, Universit\'e Pierre et Marie Curie, 98 bis Boulevard Arago, Paris 75014, France}
\affiliation{Department of Physics and Astronomy, 3701 San Martin Drive, The Johns Hopkins University, Baltimore MD 21218, USA}
\affiliation{BIPAC, 1 Keble Road, University of Oxford, Oxford OX1 3RH UK}

\begin{abstract}

We compute the sensitivity to dark matter annihilations for the forthcoming large Cherenkov Telescope Array (\CTA) in several primary channels and over a range of dark matter masses from 50 GeV up to 80 TeV. For all channels, we include  inverse Compton scattering of $e^\pm$ by dark matter annihilations on the ambient photon background, which yields substantial contributions to the overall $\gamma$-ray flux. We improve the analysis over previous work by: $i)$ implementing a spectral and morphological analysis of the $\gamma$-ray emission; $ii)$ taking into account the most up-to-date cosmic ray background obtained from a full \CTA\ Monte Carlo simulation and a description of the diffuse astrophysical emission; and $iii)$ including the systematic uncertainties in the rich observational \CTA\ datasets. We find that our spectral and morphological analysis improves the \CTA\ sensitivity by roughly a factor 2. For the hadronic channels, \CTA\ will be able to probe thermal dark matter candidates over a broad range of masses if the systematic uncertainties in the datasets will be controlled better than the percent level. For the leptonic modes, the \CTA\ sensitivity will be well below the thermal value of the annihilation cross-section. In this case, even with larger systematics, thermal dark matter candidates up to masses of a few TeV will be easily studied.

 \end{abstract}


\maketitle 

\section{Introduction}
A substantial body of astrophysical data (such as galaxy rotation curves and weak lensing observations, Large Scale Structure surveys and the precise data from CMB measurements~\cite{Bergstrom:2000pn,Bertone:2004pz}) have revealed the crucial gravitational role of a new kind of particle, dubbed as Dark Matter (DM). This provides one of the most compelling hints for beyond Standard Model (SM) physics. Nevertheless, in order to identify the microphysics nature of this particle, we need information on its mass and  interaction characteristics. One common strategy dubbed as indirect detection of DM particles aims at detecting the secondary stable SM products of a DM annihilation or decay in our Galaxy and beyond, on top of the astrophysical background. For popular DM candidates, belonging to the class of Weakly Interacting Massive Particles (WIMPs), these SM products are injected in the realm of investigation of very high energy astrophysics. In particular, the current and upcoming ground-based telescopes are very appropriate for probing the properties of WIMPs in the mass domain of few 100 GeV up to 100 TeV, by looking at the Very High Energy (VHE) $\gamma$-rays from the annihilation of such particles.

\smallskip
Over the last several years, in order to identify the nature of DM, the Imaging Atmospheric (or Air) \v{C}erenkov Telescopes (IACTs), such as \HESS~\cite{hess}, \MAGIC~\cite{magic} and \VERITAS~\cite{veritas}, have conducted rich observational programs in the inner region of the Milky Way halo~\cite{Aharonian:2006wh,Abramowski:2011hc,Abramowski:2013ax}, nearby satellite galaxies~\cite{Aharonian:2007km,Albert:2007xg,Aharonian:2008dm,Aliu:2008ny,Acciari:2010pja,2010ApJ7201174A,Abramowski:2010aa,Aleksic:2013xea,2012PhRvD85f2001A,Abramowski:2014tra} and galaxy clusters~\cite{2010ApJ710634A,Abramowski:2012au,2012ApJ757123A}. So far, the observations conducted by \HESS\ in the inner region of the Galactic halo  ($\lesssim$ 150 pc) provide the strongest constraints for TeV DM particles~\cite{Abramowski:2011hc}. This complements the bounds on the annihilation cross section in the few-tens-of-GeV mass range, coming from \FERMI\ (see e.g. those from the null observations of dwarf galaxies in $\gamma$-rays~\cite{Ackermann:2013yva}, the Galactic Centre~\cite{2011PhRvD84l3005H,2013APh4655H,Gordon:2013vta} and the diffuse $\gamma$-rays emission measurement at intermediate~\cite{Cirelli:2009dv,Papucci:2009gd,Zaharijas:2010ca} and high latitudes~\cite{Ackermann:2015tah}),  and from  the effect of generating CMB anisotropies at the recombination epoch and their evolution down to the reionization era \cite{Galli:2009zc,Slatyer:2009yq,Cirelli:2009bb,Hutsi:2011vx,Galli:2011rz}.

The next-generation of IACTs will be the forthcoming large Cherenkov Telescope Array (\CTA) that will surpass the overall performances of the present experiments. In particular, several estimates of the \CTA\ sensitivity to DM annihilation agree on the fact that there will be a substantial improvement  in the flux sensitivity compared to current IACTs up to one order of magnitude (see e.g.~Refs.~\cite{Doro:2012xx,Viana:2012zz,Wood:2013taa,Pierre:2014tra,Silverwood:2014yza}). 

\smallskip
In this paper we provide a new assessment of the \CTA\ sensitivity to DM annihilations in several primary channels  (DM DM $\rightarrow e^+e^-,  \mu^+\mu^-, \tau^+\tau^-, b\bar{b}, t\bar{t} $ and $W^+W^-$) and over a range of DM masses from 50 GeV up to 80 TeV. We improve the analysis over previous works~\cite{Doro:2012xx,Viana:2012zz,Wood:2013taa,Pierre:2014tra,Silverwood:2014yza} in several aspects. More specifically,

\begin{itemize}
\item[$\diamond$] we include for the first time in the context of the \CTA, low energy contributions of the $\gamma$-rays fluxes due to inverse Compton Scattering (ICS)  on  the ambient photon background of the $e^\pm$ from annihilating DM.  We demonstrate that this is particularly relevant for the  determination of the \CTA\ sensitivity in the leptonic channels, especially for the DM DM $\rightarrow e^+e^-$ mode. 

\item[$\diamond$] we implement a spectral and morphological analysis of the $\gamma$-rays emissions. We demonstrate that the different spectral and spatial behaviours of the DM signal compared to the background can be used to substantially improve the \CTA\ sensitivity to DM annihilation.

\item[$\diamond$] we take into account the most up-to-date Cosmic Ray (CR) background coming from a full \CTA\ Monte Carlo simulation and a description of the Galactic Diffuse Emission (GDE) estimated from \FERMI\ data. In particular, in order to maximize the impact of the GDE on the \CTA\ sensitivity, we consider an {\em isotropic} GDE coming  from the inner Galactic halo. 
With this ``extreme'' choice, we are clearly overestimating the GDE, because it is reasonable to expect that the diffuse $\gamma$-ray contamination decreases as a function of the distance from the Galactic Center. As a consequence, our results will be conservative from the point of view of  GDE  uncertainties.

\item[$\diamond$] we study the impact of the systematics errors  in the rich observational \CTA\ datasets. For example, possible sources of systematic uncertainties arise from observational issues such as different observation zenith angles, very high energy emitters and starlight gradients in the field of view, instrumental issues such as broken pixels and a non-uniform distribution of PMT quantum efficiency, and performance issues such as $\gamma$-ray and background acceptances across the field of view and normalisation between the signal and background regions~\cite{2013APh4117D}.

\end{itemize}

The rest of the paper is organized as follows: in Sec.~\ref{sec:cta} we present the main characteristics of the \CTA\ experiment; in Sec.~\ref{sec:dms} we provide a short description of the properties of the $\gamma$-rays  fluxes coming from DM annihilation, the Region of Interest (RoI) relevant for our morphological analysis and the expected number of events in the \CTA\ detector;  in Sec.~\ref{sec:bck} we quantify the CR and the GDE backgrounds; in Sec.~\ref{sec:obs} we present  our analysis methodology. In particular the implementation of our the spectral and morphological analysis in the \CTA\ likelihood and the quantification of the systematic uncertainties in the \CTA\ sensitivity; in Sec.~\ref{sec:results} we present our results and  in Sec.~\ref{sec:sum} we conclude.

\section{The Cherenkov Telescope Array}
\label{sec:cta}
\CTA\ will be the next-generation array of IACTs in VHE $\gamma$-ray astronomy. It is envisaged as a two-site array to allow full coverage of the sky. One in the northern hemisphere will have the aim of studying extragalactic sources, while the other in the Southern hemisphere will emphasize observations of the Galactic Center (GC) region since this can be observed close to the zenith during the austral winter. 

\CTA\ will consist of several tens of two or three different types of telescope displaced over a km square area, with sizes of about 6, 12 and 23 m in diameter, respectively. The sensitivity is expected to be a factor 10 better than currently operating IACTs: the field of view (FoV) of the small size telescope will be around $9^\circ$, the angular resolution roughly an arcminute, the energy resolution from about 20\% at 100 GeV to better than 5\% at 10 TeV, and a lower energy threshold of several tens of GeV~\cite{Acharya:2013sxa}. Although the final design of the array is not  settled yet, a detailed Monte Carlo study has been performed on various candidate array configurations to estimate the array performances in terms of background rejection, point-source sensitivity, angular and energy resolutions~\cite{2013APh43171B}.

\smallskip
In this study, we will consider the instrument response functions\footnote{Notice that in Ref.~\cite{Roszkowski:2014iqa} the instrument response functions are taken from unpublished materials. Therefore a direct comparison with the results presented here and in previous \CTA\  sensitivity studies is in principle not possible.} obtained for the proposed array I. This benchmark array is a balanced choice to allow for good sensitivity both in low and high energy regime. For this array, we make use of the information of the effective area for photons, residual background rate, angular and energy resolutions provided in Ref.~\cite{2013APh43171B}.

\medskip
Observations by IACTs in high energy $\gamma$-ray astronomy traditionally employ two regions on the sky expected to have roughly the same  astrophysical emission, but significantly different amounts of DM annihilation. The signal region, where the larger annihilation signal is expected to be, is usually dubbed as the ON region. The other, called the OFF region, is taken to be  larger than the ON region and is used for the background determination. The two regions are chosen in nearby regions of the sky and the statistical analysis is performed by using a test statistic defined as the difference in counts from the two regions. Observations usually cover part of the ON and OFF regions simultaneously in order to avoid additional systematics that may arise from different atmospheric, instrumental or observing conditions. 

\smallskip
In this study, we will optimize the above mentioned ON-OFF method by carrying out a full likelihood analysis which uses the expected spectral and spatial distribution of the DM signals. As will become clear later on, this improved analysis takes full advantage of the spatial morphology of the signal with respect to the background by using the available information over several degrees. Nevertheless, due to the limited \CTA\ FoV (around 7.5$^{\circ}$ for middle-size telescopes\footnote{The small and  large-size telescopes are expected to have a FoV of 9$^{\circ}$ and 4.5$^{\circ}$, respectively.}), several observational pointings  are needed to accurately map the inner 10$^{\circ}$ of the Galactic halo. In particular, due to the radial dependence of the acceptance in the FoV, one needs to develop a proper observation scan strategy to reduce the systematics  that may arise at the edges of the FoV of the observation in order to uniformly map the RoI.

In what follows, we will consider an optimized observational strategy which provides, in addition to a uniform exposure of 500 hours over all the RoIs relevant for our analysis, a substantial exposure in RoIs, used as OFF regions for background measurements, beyond the central 5 degrees. The choice of the pointing strategy to uniformly map the considered RoIs  crucially depends on the total observation time, number of pointings and grid spacing.

A general and realistic optimization of the observation strategy to search for DM in the GC region is beyond the scope of the present paper. For example, the OFF regions for core profiles (e.g.~isothermal or burkert), need to be taken far away from the ON region in order to enable a significant  gradient between the two regions. Given the fact that the \CTA\ FoV is limited, dedicated OFF observations are required to estimate the residual $\gamma$-ray background in the DM core. Such observational strategy and analysis methodology are not described here. Therefore it is worth stressing that our study, like those presented in Refs.~\cite{Doro:2012xx,Viana:2012zz, Wood:2013taa,Pierre:2014tra,Silverwood:2014yza},  is only suitable for cuspy DM profiles.

\section{Dark matter  signals}
\label{sec:dms}
\subsection{Properties of the DM fluxes}
Annihilating DM particles induce high energy $\gamma$-rays fluxes both by direct  emission ({\em prompt}) and by ICSs of $e^\pm$ produced by DM annihilation on the ambient photon background ({\em secondary}).

\begin{itemize}

\item {\em prompt} emission: The differential  $\gamma$-ray flux,  produced by  the prompt annihilation of self-conjugate DM particles of mass $\mDM$, coming from a given angular direction $\ud \Omega$, is written as  
\beq\label{promptflux}
\frac{\ud \Phi_\gamma^{\rm P}}{\ud \Omega \ud E_{\gamma}} =\frac12 \frac{r_\odot}{4\pi}\frac{\rho_\odot^2}{ \mDM^2} J(\theta) \sum_f \sigmav_f \frac{\ud N^f_\gamma}{\ud E_{\gamma}}(E_{\gamma}) \ , \qquad J(\theta)=\int_{\rm l.o.s.}\frac{\ud s}{r_\odot}\frac{\rho^2(r(s,\theta))}{\rho_\odot^2}
\eeq
where  $\sigmav_f$ and $\ud  N^f_\gamma / \ud E_{\gamma}$  are respectively the self-annihilation  cross-section and the  energy spectrum of photons per one annihilation  in the channel with final state $f$.  Here the coordinate $r$ is written $r(s, \theta)=(r_\odot^2+s^2-2 r_\odot s \cos\theta)^{1/2}$, where $s$ is the parametrization for the distance along the line-of-sight (l.o.s.), $\theta$ is the aperture between the direction of observation and the Galactic plane and $r_\odot=8.33$ kpc is the Sun's location with respect to  the GC. As usual, the aperture $\theta$ in polar coordinates can be expressed in terms of galactic latitude $b$ and longitude $\ell$  via the relation $\theta(b,\ell)=\arccos\left(\cos b \cos \ell\right)$. The function $J(\theta)$, commonly referred as to the {\it J-factor}, integrates  the square of the DM density $\rho$ along the  line of sight. For the computation of the {\it J-factor}, we will always assume an Einasto profile 
\beq
\rho(r) = \rho_s  \exp \left[-\frac{2}{\alpha_s}\left(\Big(\frac{r}{r_s}\Big)^{\alpha_s }-1\right)\right] \ , 
\eeq
whose parameters ($\rho_s=0.033 \mbox{ GeV}/\mbox{cm}^3, r_s=28.44 \mbox{ kpc}, \alpha_s= 0.17 $) are taken by demanding that: $i)$ the DM  density at the Sun location is $\rho_\odot=0.3$ GeV/cm$^3$; $ii)$ the total DM mass contained in 60 kpc is $M_{60}=4.7 \times 10^{11}M_\odot$~\cite{Xue:2008se}.  In order to derive the prompt $\gamma$-rays fluxes, we will compute $J(\theta)$ and $\ud  N^f_\gamma / \ud E_{\gamma}$, by using the tools in Ref.~\cite{Cirelli:2010xx}. 

It is worth stressing here that either other parametrizations  or parameters (e.g.~the authors of Ref.~\cite{Silverwood:2014yza} quote $\rho_\odot=0.4$ GeV/cm$^3$ at $r_\odot=8.5$ kpc) of the Einasto profile can yield a substantially different value of the $J$-factor. Therefore, when comparing different results,  we have always to  be  aware about this fine distinction. In particular, our present choice of parameters leads to the smallest value of the $J$-factor in the inner Galactic halo compared to previous analyses~\cite{Doro:2012xx,Viana:2012zz, Wood:2013taa,Pierre:2014tra,Silverwood:2014yza}. As a consequence, our results, from the point of view of DM profile uncertainties, are the most conservative.

\item {\em secondary} emission: The differential  $\gamma$-ray flux   produced by  the IC radiative processes within an angular region $\ud \Omega$ of the sky can be obtained by convolving the number density in the emitting medium  with the differential power that it radiates. We write 
\beq\label{ICSflux}
\frac{\ud \Phi_\gamma^{\rm IC}}{\ud \Omega \ud E_{\gamma}} =\frac1{4\pi  E_\gamma} \int_{\rm l.o.s.} \hspace{-.25cm} \ud s \, 2 \int_{m_e}^{\mDM} \ud E_e \,  \mathcal P_{\rm IC}(E_\gamma,E_e, r) \frac{\ud n_{e^\pm}}{\ud E_e}(E_e, r ) \ ,
\eeq
where the factor 2 takes into account that an equal population of electrons and positrons is produced by DM annihilations. In Eq.~\eqref{ICSflux}, $\ud n_{\rm e^\pm}/\ud E_e$ is the number density of electrons (or positrons) after energy losses and diffusion which obeys the diffusion-loss equation given for instance in Refs.~\cite{Salati:2007zz, Cirelli:2010xx}, 
while $\mathcal P_{\rm IC}=\sum_i \mathcal P_{\rm IC}^i$ is the total differential power radiated into photons by the ICS mechanism\footnote{The sum runs over the three different species of the ambient photon background; i.e.~starlight, IR and CMB. For starlight and IR light, we extract the maps of their distribution from Galprop \cite{Vladimirov:2010aq}.}. We refer the reader to Refs.~\cite{Cirelli:2009vg,Cirelli:2010xx} 
and references therein where the analytic formula of $\mathcal P_{\rm IC}$, valid in the full Klein-Nishina case, and a semi-analytic solution of $\ud n_{\rm e^\pm}/\ud E_e$ are provided. 

Following Ref.~\cite{Cirelli:2010xx}, Eq.~\eqref{ICSflux} can be rewritten as a convolution of the $e^\pm$ injection spectrum $\ud N_{e^\pm}^f/\ud E_e$ in the channel with final state $f$ with a halo function $I_{\rm IC}$ for the IC radiative process. We explicitly write
\beq\label{ICSfluxPPPC}
\frac{\ud \Phi_\gamma^{\rm IC}}{\ud \Omega \ud E_{\gamma}} = \frac12 \frac{r_\odot}{4\pi}\frac{\rho_\odot^2}{ \mDM^2}\frac1{E_\gamma^2}\int_{m_e}^{\mDM} \ud E_s \, I_{\rm IC}(E_\gamma,E_s,b,\ell) \,  \sum_f \sigmav_f \frac{\ud N^f_{e^\pm}}{\ud E_s}(E_s) \ ,
\eeq
where $E_s$ is the electron (or positron) injection energy. In order to derive the differential $\gamma$-ray flux produced by the secondary emission, we will compute $I_{\rm IC}$ and $\ud  N^f_{e^\pm} / \ud E_s$,  by using the tools in Ref.~\cite{Cirelli:2010xx}
and then we will numerically perform the integral on the right-handed side of Eq.~\eqref{ICSfluxPPPC}.

\end{itemize}

\subsection{Definition of the RoIs relevant for our analysis}
\label{RoIsdef}

\begin{table}
\centering
\begin{tabular}{c|c|c|c|c}
$i$-esime RoI & Angular size ($\Delta\Omega_i$) in [sr]&  \multicolumn{3}{c}{$J$-factor ($\bar J^{\Delta\Omega}_i \rho_\odot^2 r_\odot$) in [GeV$^2$/cm$^5$]}  \\
\hline
&  & Einasto used here & Einasto as in~\cite{Silverwood:2014yza} & NFW as in~\cite{Cirelli:2010xx}  \\
\hline
\hline
first RoI ($\bar \theta_i=1^\circ$) & $5.91 \times 10^{-4}$ & $1.42 \times 10^{21}$ & $4.61 \times 10^{21}$ & $1.09 \times 10^{21}$ \\
second RoI ($\bar \theta_i=2^\circ$) & $2.51 \times 10^{-3}$  & $3.17 \times 10^{21}$ & $9.56 \times 10^{21} $ & $2.02 \times 10^{21}$ \\
third RoI ($\bar \theta_i=3^\circ$) &  $4.42 \times 10^{-3}$  & $3.37 \times 10^{21}$  & $9.59 \times 10^{21} $ & $2.03 \times 10^{21}$ \\
fourth RoI ($\bar \theta_i=4^\circ$) &  $6.33 \times 10^{-3}$  & $3.30 \times 10^{21}$  & $9.02 \times 10^{21} $ & $1.98 \times 10^{21}$ \\
fifth RoI ($\bar \theta_i=5^\circ$) &  $8.23 \times 10^{-3}$  & $3.15 \times 10^{21}$  & $8.35 \times 10^{21} $ & $1.92 \times 10^{21}$ \\
\hline
\hline
\end{tabular}
\caption{ \small \label{tableJ} Angular size of the $i$-esime RoI and corresponding value of the $J$-factor in units of GeV$^2$/cm$^5$ for the Einsato profile considered here. For a sake of comparison, averaged $J$-factors are also given for an alternative normalisation of the Einasto profile~\cite{Silverwood:2014yza} and a NFW profile~\cite{Cirelli:2010xx}.}
\end{table}

\begin{figure}[t]
\centering
\includegraphics[width=.32\textwidth]{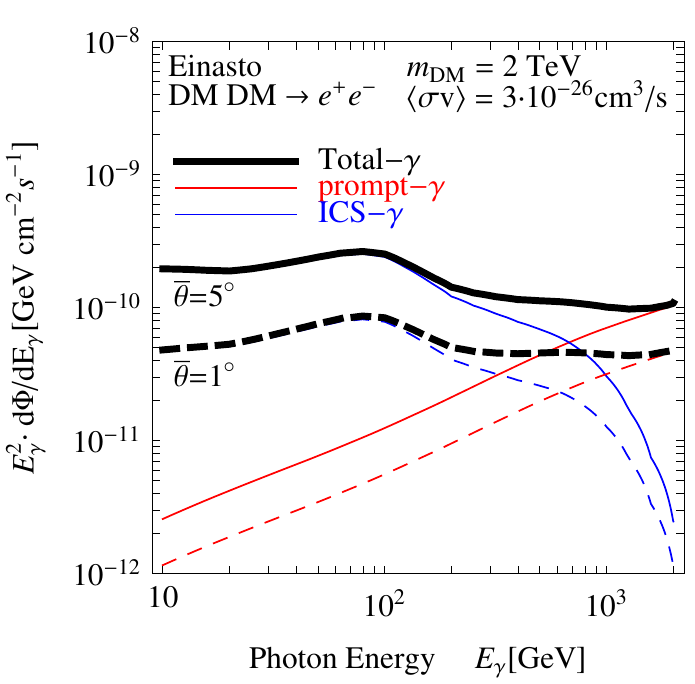}
\includegraphics[width=.32\textwidth]{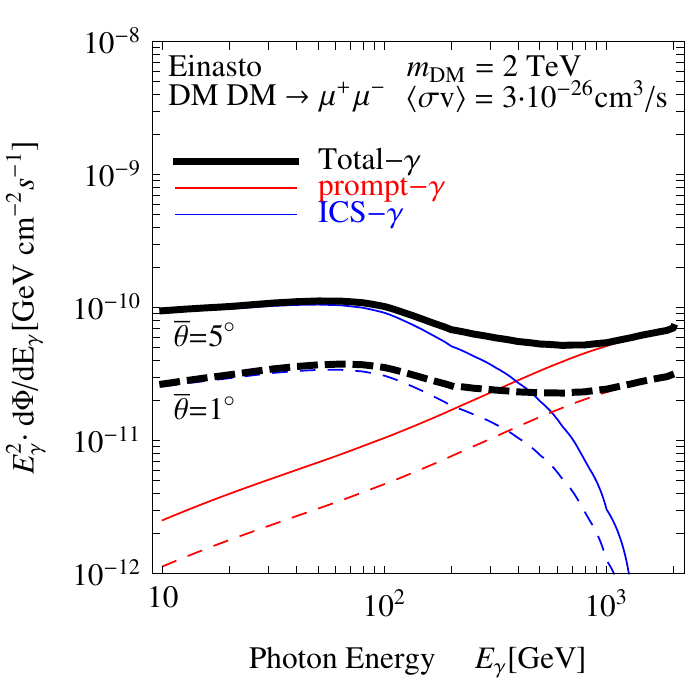}
\includegraphics[width=.32\textwidth]{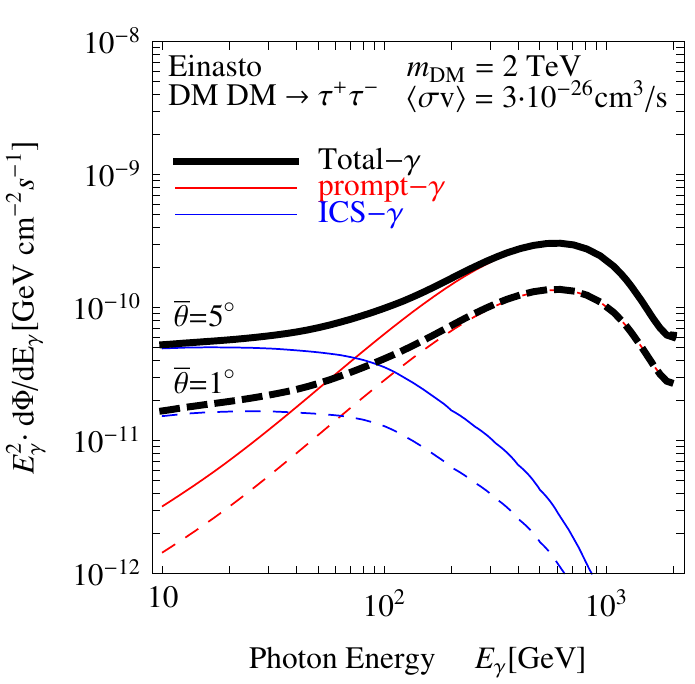}

\includegraphics[width=.32\textwidth]{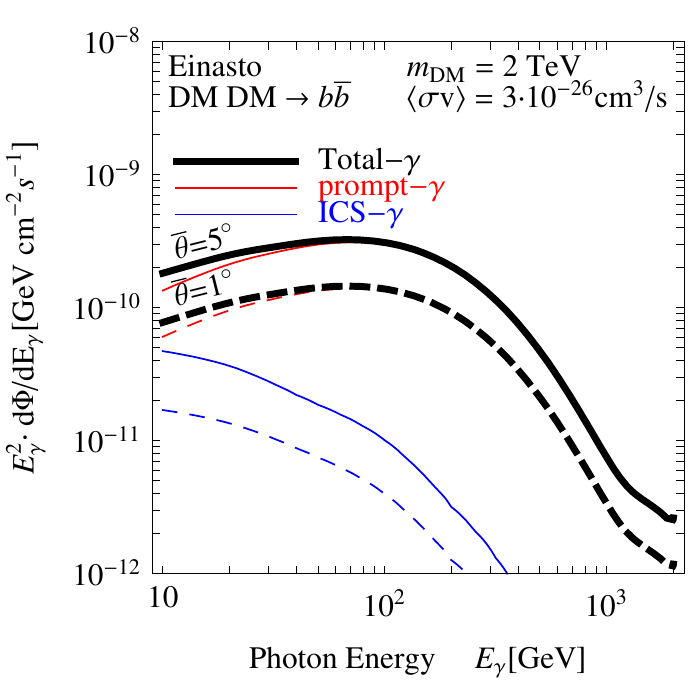}
\includegraphics[width=.32\textwidth]{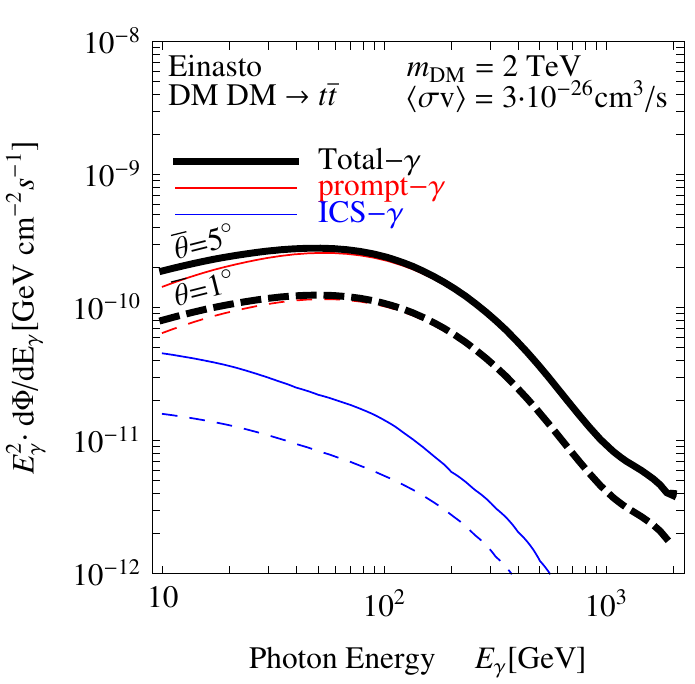}
\includegraphics[width=.32\textwidth]{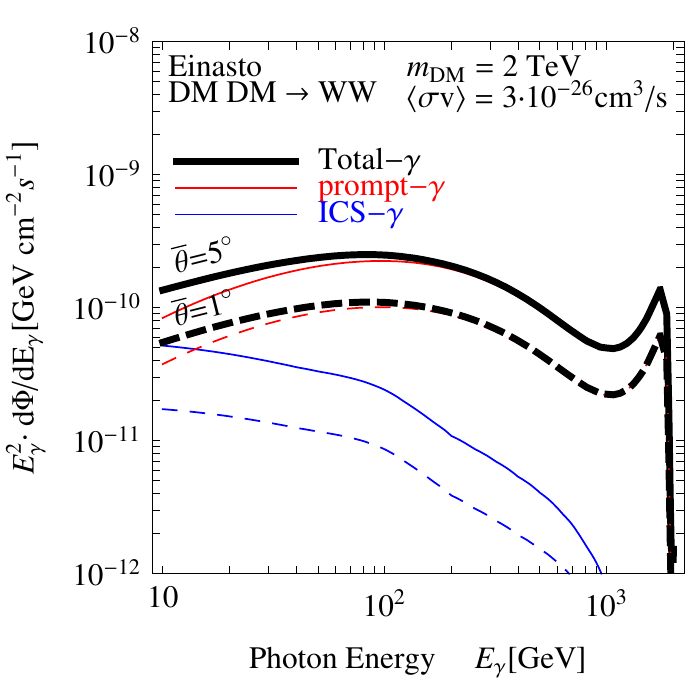}
\caption{\small Several example of $\gamma$-rays fluxes for 2 TeV DM candidates annihilating with thermal cross section ($\rm \langle \sigma v \rangle  = 3 \times 10^{-26} cm^3/s$)  into leptonic channels (first row) and hadronic ones (second row). In all plots, the different  hashing shows the predicted signal in two  benchmark RoIs with apertures $\bar\theta_1$ (dotted line), 
and $\bar\theta_5$ (solid line). The red, blue and thick black lines denote instead the  spectral features of the prompt, ICSs and total fluxes respectively. See the text for further details.}
\label{fig:En2_exp}
\end{figure}

Having at our disposal the fluxes per steradian, one needs to integrate Eqs.~(\ref{promptflux}, \ref{ICSfluxPPPC}) over a region $\Delta\Omega$ which cover, for instance,  a given \CTA\ observational window. In particular, we will consider five RoIs corresponding to annuli
centered in the Galactic Center with $i$-esime aperture $\bar\theta_i$ and constant thickness $\Delta\theta$, 
minus  rectangular regions with latitude within $b_{\rm min}=0^\circ \leq b \leq b_{\rm max}=0.3^\circ$ and  $i$-esime longitude which intercepts the annulus thickness\footnote{In the central 300 pc of the GC ($\ell \simeq 0^\circ$) and at Galactic latitudes with $|b|<0.3^\circ$, the \HESS\  collaboration discovered two bright sources: $i)$ HESS J1745-290 coincident in position with the supermassive black hole Sgr A$^*$~\cite{Aharonian:2004wa}; and $ii)$ HESS J1747-281 coincident with the supernova/pulsar wind nebula G0.9+0.1~\cite{Aharonian:2005br},  together with a strong astrophysical diffuse emission~\cite{Aharonian:2006au} (see Sec.~IV.B). This region is excluded in order to avoid VHE $\gamma$-ray background contamination in the considered RoIs.}.
The solid angle $\Delta\Omega_i$ covered by the $i$-esime RoI will then be
\beq\label{CTARoIs}
\Delta\Omega_i=\Delta\Omega^{\rm ann}_i-\Delta\Omega^{\rm rec}_i, \qquad \left\{\begin{array}{l}
\displaystyle \Delta\Omega^{\rm ann}_i= 2\pi\int_{\bar\theta_i-\Delta\theta}^{\bar\theta_i} \ud \theta  \sin\theta  \qquad \text{($i$-esime annulus)} \\
\displaystyle \Delta\Omega^{\rm rec}_i= 4\int_{\bar\theta_i-\Delta\theta}^{\bar\theta_i} \ud \ell  \int_{b_{\rm min}}^{b_{\rm max}} \ud b  \cos b \qquad \text{($i$-esime rectangular region)}
\end{array}\right. \ ,
\eeq
and the corresponding  averaged {\it J-factor} $\bar J^{\Delta\Omega}_i=\bar J^{\Delta\Omega^{\rm ann}}_i-\bar J^{\Delta\Omega^{\rm rec}}_i$ and averaged {\it IC halo function} $\bar I_{{\rm IC},  i}^{\Delta\Omega}(E_\gamma,E_s)=\bar I_{{\rm IC},  i}^{\Delta\Omega^{\rm ann}}(E_\gamma,E_s)-\bar I_{{\rm IC},  i}^{\Delta\Omega^{\rm rec}}(E_\gamma,E_s)$ are obtained by integrating the function $J(\theta(b,\ell))$ and $I_{\rm IC}(E_\gamma,E_s, b, \ell)$ over the region $\Delta\Omega_i$ defined in Eq.~(\ref{CTARoIs}). For our morphological analysis, the annulus thickness  is always $\Delta \theta = 1^\circ$ and we choose five values of $\bar\theta_i$ ranging from $1^\circ$ to $5^\circ$ with step of $1^\circ$. In the rest of the paper the label $i=1,...,5$ will always refer to RoIs with  $\bar \theta_i=1^\circ,...,5^\circ$. 

Table~\ref{tableJ} shows the angular size of the $i$-esime RoI together with the corresponding value of the averaged $J$-factor in units of GeV$^2$/cm$^5$. As one can see,   although the angular size of the outer RoIs is bigger, the values of the $J$-factor stay almost constant for regions above the second RoI.  Averaged $J$-factors, assuming an Einasto profile with the same parameters adopted in Ref.~\cite{Silverwood:2014yza} and a NFW profile as in Ref.~\cite{Cirelli:2010xx}, are showed in the same table for a sake of comparison. As one can see, the Einasto profile used in this study provides conservative estimate of the $J$-factor in the considered RoIs with respect to the same profile but with different parameters adopted in Ref.~\cite{Silverwood:2014yza}. On the other hand, the NFW profile, which is cuspier at the GC than Einasto, provides a shightly smaller value of the $J$-factor in the considered RoI. This is mainly due to the fact that the Einasto profile is somewhat more chubby than NFW at few 100 pc to kpc scales (see e.g.~Fig.~1 of Ref.~\cite{Cirelli:2010xx}).

\smallskip
The integrated total flux (primary + secondary) coming from the $i$-esime RoI will finally write  
\beq\label{TotalFlux}
\frac{\ud \Phi_{\gamma, i}^{\rm tot}}{ \ud E_{\gamma}}= \frac{\ud \Phi_{\gamma, i}^{\rm P}}{ \ud E_{\gamma}}+\frac{\ud \Phi_{\gamma, i}^{\rm IC}}{ \ud E_{\gamma}}, \qquad  \left\{\begin{array}{l}
\displaystyle \frac{\ud \Phi_{\gamma, i}^{\rm P}}{ \ud E_{\gamma}} = \frac12 \frac{r_\odot}{4\pi}\frac{\rho_\odot^2}{ \mDM^2} \bar J^{\Delta\Omega}_i \sum_f \sigmav_f \frac{\ud N^f_\gamma}{\ud E_{\gamma}}(E_{\gamma}) \  \\
\displaystyle \frac{\ud \Phi_{\gamma, i}^{\rm IC}}{ \ud E_{\gamma}} =  \frac12 \frac{r_\odot}{4\pi}\frac{\rho_\odot^2}{ \mDM^2}\frac1{E_\gamma^2}\int_{m_e}^{\mDM} \ud E_s \, \bar I_{{\rm IC},  i}^{\Delta\Omega}(E_\gamma,E_s) \,  \sum_f \sigmav_f \frac{\ud N^f_{e^\pm}}{\ud E_s}(E_s) \ 
\end{array}\right. \ .
\eeq
In Fig.~\ref{fig:En2_exp}, we show the $\gamma$-ray fluxes  for DM candidates with thermal cross-section ($\sigmav_{\rm th} = 3\times 10^{-26}$ cm$^3/$s) in leptonic channels  (first row), and in hadronic ones (second row). In all plots the  hashing shows the predicted signal in two  benchmark RoIs with apertures $\bar\theta_1$ (dotted line), 
and $\bar\theta_5$ (solid line). Together with the values reported in Tab.~\ref{tableJ}, we can clearly see that the fluxes increase of roughly a factor 3 moving from the closest to the farthest RoIs used in our morphological analysis. 

As an aside, the different plots in Fig.~\ref{fig:En2_exp}, allow us also to appreciate the different spectral features of the primary  (red lines), secondary (blue lines) and total (thick black lines) emissions. As we can see,  for a 2 TeV  DM annihilating into $e^+e^-$ and $\mu^+\mu^-$, the $\gamma$-rays flux spectra are dominated by the  secondary emission up to a  photon energy close to the DM mass. For the $\tau^+\tau^-$ channel, the prompt emission is more pronounced and it dominates above a photon energy of roughly 100 GeV, while for the hadronic scenarios the secondary emissions are always subdominant in the entire \CTA\  energy window. 

Finally it is worth noticing that for the channel DM DM$\rightarrow W^+W^-$, a pronounced spectral feature close to  $\mDM$ appears. This is due to the fact that the photons coming from the splitting $W^\pm \rightarrow W^\pm \gamma$ contribute to the signal in a significant way. That electroweak process in fact, has two ``soft'' singularities: one comes from the usual soft photon, the other from a soft $W$ (thus a $\gamma$ that carries away almost all the $W$ energy). The latter divergence is screened by the mass of the $W$. Hence, we expect that the feature one can observe in the $\gamma$-ray fluxes from the $W^+W^-$ channel, is only relevant for very heavy DM candidates ($\mDM \gg m_W$). We stress here, that this pronounced spectral feature can be used for deriving constraints for lines-like searches has already done for the DM DM$\rightarrow \gamma \gamma$ or $\gamma Z$ by the \HESS\ collaboration~\cite{Abramowski:2013ax}.

\subsection{Expected photon counts in the CTA array}
\begin{figure}[t]
\centering
\includegraphics[width=.325\textwidth]{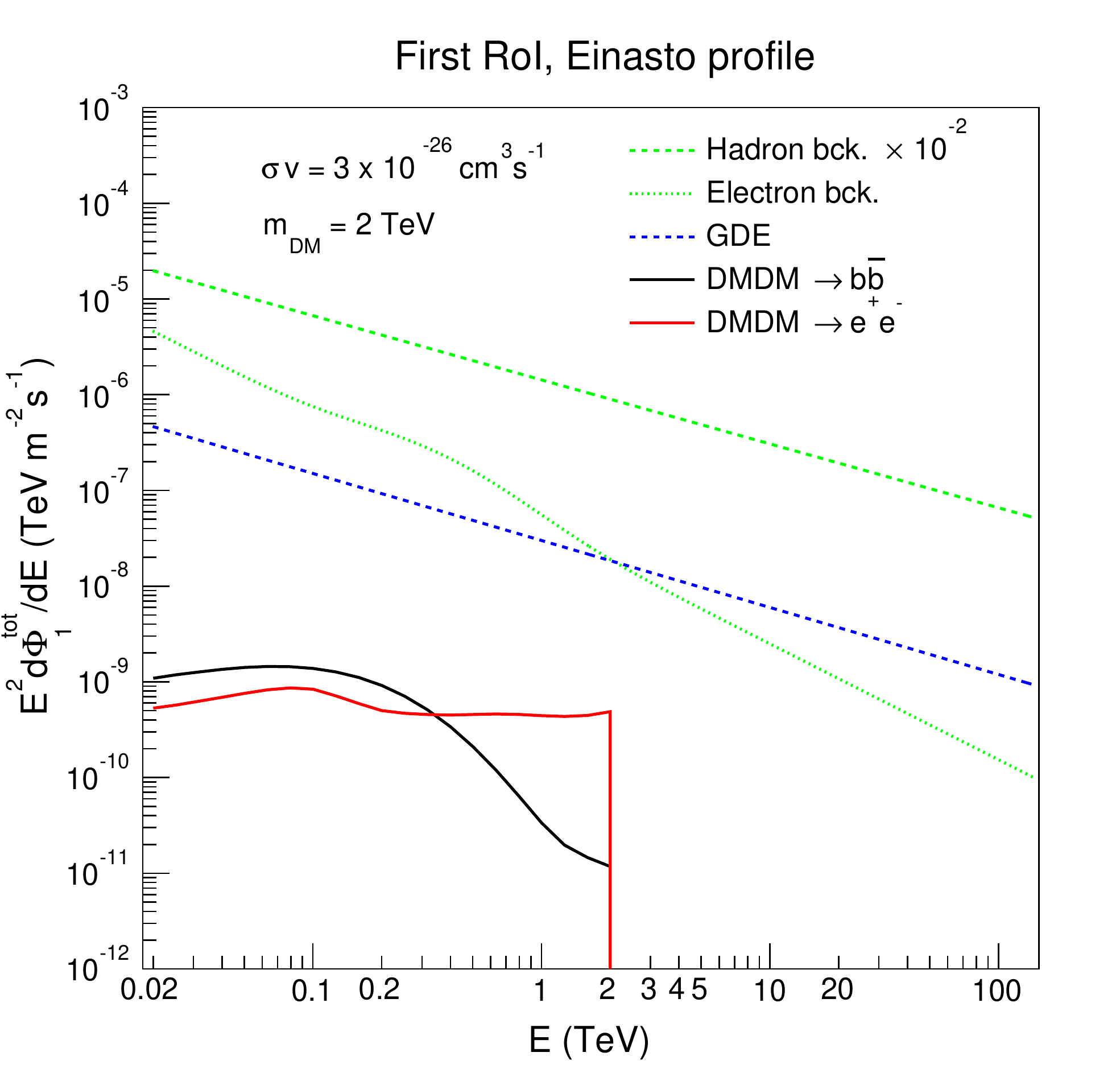}\
\includegraphics[width=.325\textwidth]{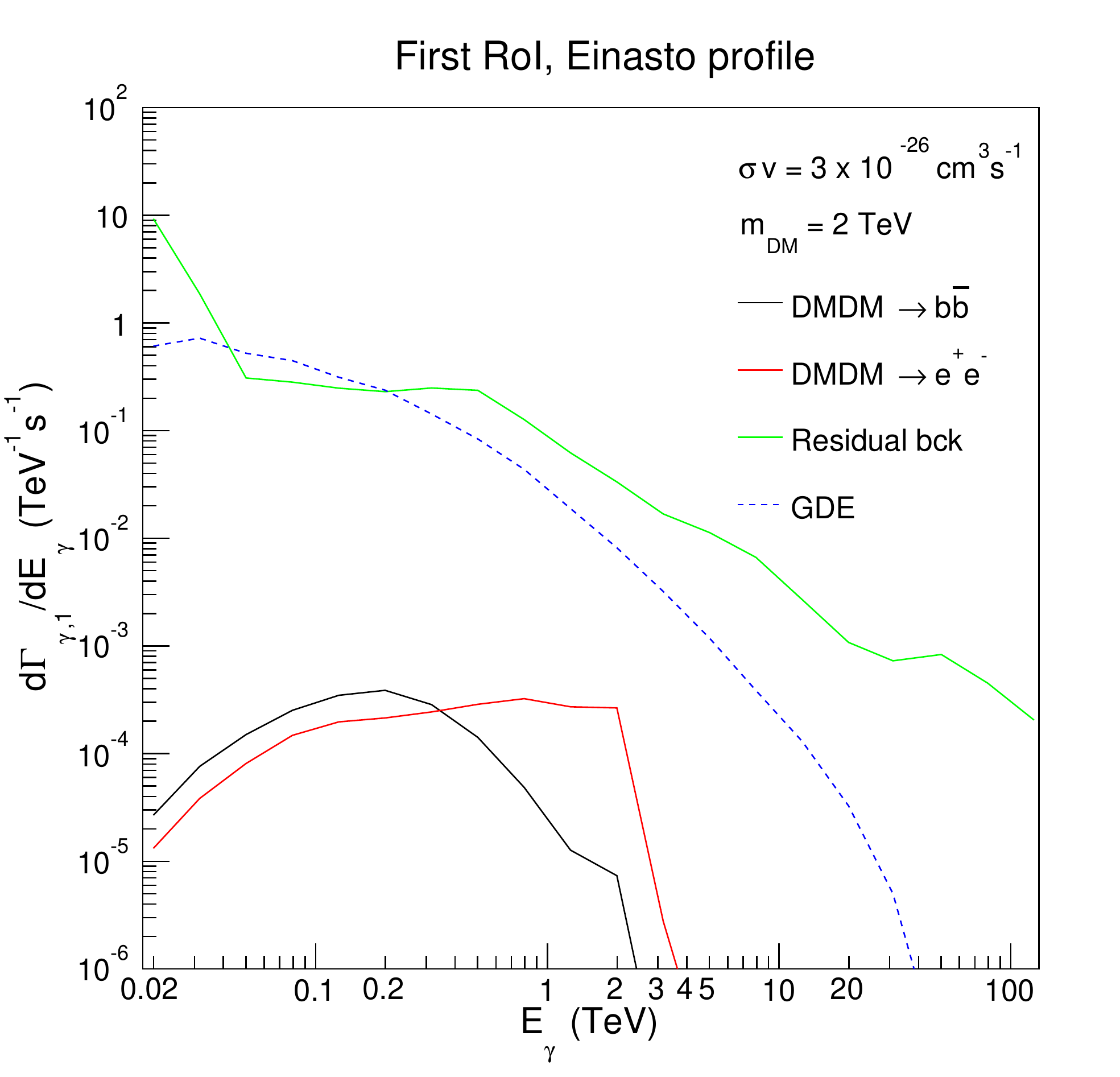}\
\includegraphics[width=.325\textwidth]{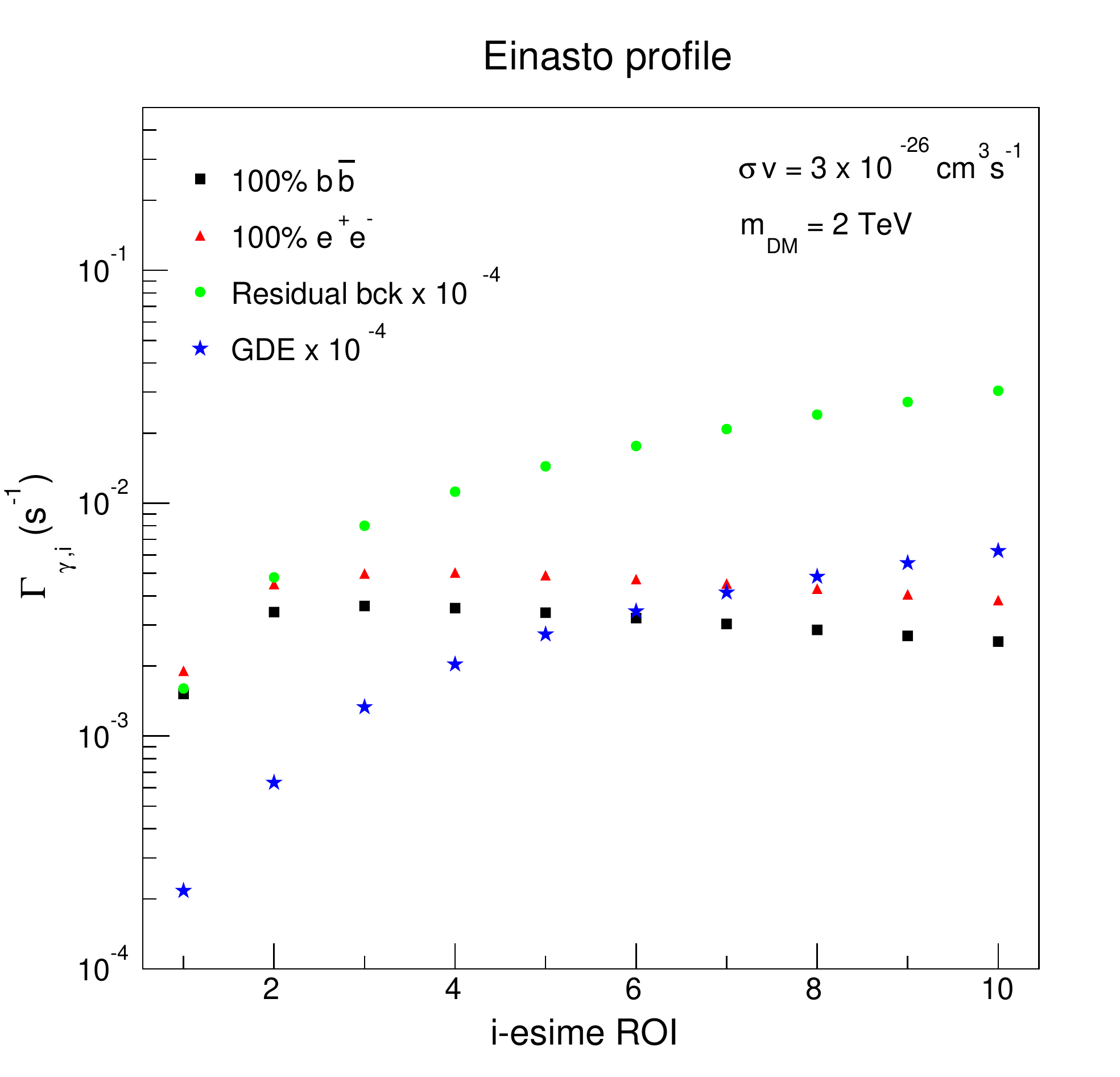} 

\caption{\small Spectral and spatial morphology of  typical DM signals together with the irreducible backgrounds. 
{\em Left panel:} Expected DM differential fluxes in the first RoI, considering 2 TeV thermal DM candidates annihilating into $b \bar b$ (black solid line) and  $e^+e^-$ (red solid line) primary channels. For a sake of illustration, we show a spectrum for the CR hadrons (protons + nuclei) background (dashed green line) multiplied by a rejection factor $\epsilon=10^{-2}$ and a log-normal distribution for the electron spectrum (green dotted line) without rejection factor ($\epsilon_e=1$). The parametrizations of the irreducible backgrounds are given in Tab.~3 of Ref.~\cite{2013APh43171B}. Notice that the energy $E$ in the $y$-axis, refers to the ``true'' energy of the particle which initiates the shower. The GDE flux detected by \FERMI\ extrapolated at high energy is also shown as dashed blue line. {\em Central panel:} Expected differential count rate in the first RoI considering  2 TeV DM candidates with thermal cross section into $b \bar b$ (black solid line) and  $e^+e^-$ (red solid line) primary channels. The most up-to-date CR background from a \CTA\ Monte Carlo simulation (solid green line) and the GDE (dashed blue line) detected by \FERMI\  in the same region are also presented for the sake of comparison.
{\em Right panel:} Expected  total  rate summed  over all energy bins in the $i$-esime RoI and for the same benchmark DM candidates (red filled triangle for annihilating DM into $e^+e^-$ pairs, black filled square for $b\bar b$ primary channel).  The irreducible CR hadrons background in the $i$-esime RoI is identified as green filled circles, while our ``extreme''  {\em isotropic} GDE as blue-filled stars. See the text for further details.
}
\label{fig:plot_S_B}
\end{figure}

In order to properly reproduce the expected photon counts in the \CTA\ array we need to take into account the characteristics and performances of the detector.  As stated in Sec.~\ref{sec:cta} we  consider the instrument response functions obtained for the proposed array I. For a given observation time $T_{\rm obs}$, the  number of observed $\gamma$-ray events in the energy bins of the benchmark array is then obtained by convolving Eq.~(\ref{TotalFlux}) with the energy-dependent effective area for photon $\mathcal A^\gamma_{\rm eff}(E_\gamma)$ and the Gaussian energy response.  In the $i$-esime RoI and $j$-esime energy bin with width $\Delta E_j$, it  writes
\beq
N_{\gamma, ij}^{\rm S}=T_{\rm obs} \int_{\Delta E_j}  \hspace{-.15cm}  \ud E_\gamma \, \frac{\ud \Gamma_{\gamma, i}^{\rm S}}{\ud E_\gamma} \ , \qquad \qquad  \frac{\ud \Gamma_{\gamma, i}^{\rm S}}{\ud E_\gamma}= \int_{-\infty}^{+\infty} \hspace{-.15cm} \ud E_\gamma' \, \frac{\ud \Phi_{\gamma, i}^{\rm tot}}{ \ud E_{\gamma}}(E_\gamma') \, \mathcal A^{\gamma}_{\rm eff}(E_\gamma') \, \frac{e^{-\frac{(E_\gamma-E_\gamma')^2}{2\sigma^2(E_\gamma')}}}{\sqrt{2\pi\sigma^2(E_\gamma')}}  \ ,
\eeq
where $\sigma(E_\gamma) =  \delta_{\rm res}(E_\gamma)  / \sqrt{8 \ln(2)}$ is the energy dependent Gaussian width and $\delta_{\rm res}(E_\gamma)$ is the energy resolution taken from Ref.~\cite{2013APh43171B}. Here $\ud \Gamma_{\gamma, i}^{\rm S}/\ud E_\gamma$ is the expected  differential rate of counts. The left and central panels of Fig.~\ref{fig:plot_S_B} shows the spectral morphology of $E_\gamma^2 \cdot \ud \Phi_{\gamma, 1}^{\rm tot}/\ud E_\gamma$ and  $\ud \Gamma_{\gamma, 1}^{\rm S}/\ud E_\gamma$ respectively, considering  2 TeV DM candidates with thermal cross section into $b \bar b$ (black solid line) and  $e^+e^-$ (red solid line) primary channels. From the comparison between the left and central panels, we can see that, at low energy, the DM signals are suppressed due to the reduced \CTA\ effective area. This effect is sizeable below roughly 100 GeV: indeed it can reduce the low-energy rate by about 2 orders of magnitude.  On the other hand, at high energy, the rates do not go to zero exactly at $m_{\rm DM}=2$ TeV. This is because of the finite energy resolution of the detector that subtracts part of the signal close to $m_{\rm DM}$ and redistributes it towards higher energy.

\smallskip
The right panel of Fig.~\ref{fig:plot_S_B}, shows instead the spatial morphology of the total rates summed over all the energy bins above 30 GeV ($\Gamma_{\gamma, i}^{\rm S}=1/T_{\rm obs} \sum_j N_{\gamma, ij}^{\rm S}$) in the $i$-esime RoI and for the same benchmark DM candidates (green filled triangle for annihilating DM into $e^+e^-$ pairs, red filled square for $b\bar b$ primary channel). It is worth noticing  a sign change of the slope of the DM signal shape between the RoIs 3 and 4 (see also Tab.~\ref{tableJ}). We anticipate that the gentle spatial dependence of the DM signal with respect to the steeper one of the irreducible backgrounds (right panel of Fig.~\ref{fig:plot_S_B}) will allow for a substantial discrimination power that can be used to improve the \CTA\ sensitivity.

\section{CTA irreducible backgrounds}
\label{sec:bck}
In case of IACTs, the dominant backgrounds for DM searches at the Galactic Center are 
the residual flux of CR hadrons and electrons and the GDE. In what follows, we discuss and quantify  
these irreducible backgrounds that erroneously pass the photon hardware trigger and analysis cuts. 

\subsection{CR protons and electrons background events}
The numerous interactions between CR hadrons (protons + nuclei) and the Earth's atmosphere trigger hadronic showers which induce electromagnetic sub-cascades due to the decay of  neutral pions into photons. The resulting electromagnetic cascade has a very different spatial morphology with  respect to the one originated by CR electrons and photons, making possible a shower shape discrimination. Nevertheless, since the flux of CR  hadrons is much larger than the one of CR electrons and photons, an electromagnetic cascade from a pion decay can be erroneously reconstructed as a $\gamma$-ray if the rejection factor of the instrument is not very small. For a sake of illustration, the dashed green line in the left panel of Fig.~\ref{fig:plot_S_B} shows the spectrum of the isotropic CR hadrons background taken from Ref.~\cite{2013APh43171B} in the first RoI, assuming a rejection factor $\epsilon=10^{-2}$ as done in Ref.~\cite{Silverwood:2014yza}. As will become clear later on, assuming a constant $\epsilon$ is not justified since it is energy dependent. Furthermore, it is also important to point out that the hadrons flux in the left panel of Fig.~\ref{fig:plot_S_B}, is shown as a function of the ``true'' energy of the particle which initiate the cascade. This is not the measured photon energy by \CTA\ because of the reduced Cherenkov light emitted by the hadronic shower. Hence, an estimation of the CR hadrons background in the \CTA\ energy window, can be obtained by shifting down the energy of the hadron. For example, assuming that only protons contribute to the CR hadrons flux, one can shift $E_p$ to lower energies by a factor of 3, as done in Ref.~\cite{Silverwood:2014yza} (i.e.~one can multiply the differential hadrons flux in the left panel of Fig.~\ref{fig:plot_S_B} by a factor of  $3\cdot 3^{-2.7} \simeq 0.155$), because among all the pions produced by the proton interactions with the Earth's atmosphere, only 1/3 are neutral and in turn initiate electromagnetic sub-showers.  Nevertheless, this rescaling is a rough approximation because also heavier species, especially He, are relevant (see e.g.~Tab.~3 of Ref.~\cite{2013APh43171B}).  Therefore, we stress here that the left panel of Fig.~\ref{fig:plot_S_B} can be only used for rough comparison. 

CR electrons constitute an annoying background for $\gamma$-ray observations at Earth's surface, because they initiate quasi indistinguishable electromagnetic cascades with respect to those originated by $\gamma$-ray interactions. However, some discrimination between electrons and $\gamma$-rays is possible by reconstruction of the primary interaction depth of the event~\cite{2013APh43171B}. For a sake of illustration, the dotted green line in the left panel of Fig.~\ref{fig:plot_S_B} shows the spectrum of the isotropic CR electrons background taken from Ref.~\cite{2013APh43171B} in the first RoI, without rejection factor ($\epsilon_e=1$).

\smallskip
The total number of residual background events (hadrons + electrons) is then given by: 
\beq
N_{\gamma, ij}^{\rm CR}= T_{\rm obs} \int_{\Delta E_j}  \hspace{-.15cm}  \ud E_\gamma \, \frac{\ud \Gamma_{\gamma, i}^{\rm CR}}{\ud E_\gamma} \ , \qquad \qquad  \frac{\ud \Gamma_{\gamma, i}^{\rm CR}}{\ud E_\gamma}= \int_{-\infty}^{+\infty} \hspace{-.15cm} \ud E_\gamma' \, \frac{\ud \Phi_{\gamma}^{\rm CR}}{ \ud E_{\gamma}\ud \Omega}(E_\gamma') \Delta\Omega_i \, \mathcal A^{\rm CR}_{\rm eff}(E_\gamma') \, \frac{e^{-\frac{(E_\gamma-E_\gamma')^2}{2\sigma^2(E_\gamma')}}}{\sqrt{2\pi\sigma^2(E_\gamma')}}  \ ,
\eeq
where  $\ud \Phi^{\rm CR}_{\gamma}/(\ud E_{\gamma}\ud \Omega)$ is the  total CR  background flux per steradian and $\mathcal A^{\rm CR}_{\rm eff}(E_\gamma)$ is the energy-dependent effective area for CR. In our analysis, the counting rate of the irreducible background events $\ud \Gamma_{\gamma, i}^{\rm CR}/\ud E_\gamma$ is directly extracted  from Ref.~\cite{2013APh43171B} and  provides the most up-to-date background computation for \CTA. In so doing, we neither assume the rejection factors for CR hadrons and electrons, nor the shift at low energy to take into account for the reduced Cherenkov light emitted by hadronic showers, because they directly come from a full \CTA\ Monte Carlo simulation. 
The solid green line in the central panel of Fig.~\ref{fig:plot_S_B} shows the spectral morphology of this rate in the first RoI for the considered array configuration. 

\smallskip
We have checked that our irreducible background, which comes from  full \CTA\ Monte Carlo simulations, is different compared to the analytic one implemented in Ref.~\cite{Silverwood:2014yza}. To quantify the discrepancies, we can multiply the counting rate of the irreducible background events in the central panel of Fig.~\ref{fig:plot_S_B} by the factor $E_\gamma^2 /(\mathcal A^{\rm CR}_{\rm eff}(E_\gamma)\Delta\Omega_1)$ and directly compare it with the solid black line in Fig.~2 of \cite{Silverwood:2014yza}. We get a higher total background especially above 1 TeV where the difference can be bigger than one order of magnitude. 

This improved background treatment will then degrade the sensitivity to DM annihilations compared to Ref.~\cite{Silverwood:2014yza}.  
We think that such differences are due to three main reasons. First, the latter paper tried to model the irreducible background by introducing a constant rejection factor by hand on the incoming proton flux (in particular in Ref.~\cite{Silverwood:2014yza}, $\epsilon_p=10^{-2}$). This is a rough approximation because generally $\epsilon_p$ depends on the energy.  Second, they used the same effective areas for CRs and photons. This is  an  unjustified assumption, because the effective area depends on the particle which initiate the shower. Third, they assume that only protons are relevant. This is again not true because also the CR He interactions with the Earth's atmosphere generate an important irreducible background in the \CTA\ energy window. Hence, taking a constant efficiency, assuming that only protons are relevant and taking $\mathcal A^{\rm CR}_{\rm eff}(E_\gamma) \equiv \mathcal A^{\gamma}_{\rm eff}(E_\gamma)$, can lead to underestimate the total CR background.

 \smallskip
As an aside, the central panel of Fig.~\ref{fig:plot_S_B} allows us also to appreciate  the potential discrimination power from the peculiar spectral shape of the DM signal, characterised by either sharp energy cut-off or bump-like features, with respect to the smoother spectral shape of the residual background. The left and central panels of Fig.~\ref{fig:plot_S_B} also shows the GDE $\gamma$-rays (dashed blue line) background, which we discuss in detail in the next section. 
 
 \smallskip
The right panel of the same figure shows instead the spatial morphology of the total rates of background events summed over all the energy bins above 30 GeV ($\Gamma_{\gamma, i}^{\rm CR}=1/T_{\rm obs} \sum_j N_{\gamma, ij}^{\rm CR}$). As opposed to the DM signal, the background shape over the RoIs (green filled circle) is monotonic. This is due to the fact that the background is roughly isotropic and therefore the expected background events increase proportionally with the size of the RoIs ($N_{\gamma, ij}^{\rm CR}/N_{\gamma, kj}^{\rm CR}=\Delta\Omega_i/\Delta\Omega_k$). From the different energy and spatial behaviours compared to the DM signals, we then foresee a substantial gain in sensitivity.

\subsection{GDE background events}
\label{sec:GDE}

The inner Galactic halo is a very crowded region with numerous astrophysical emitters in the VHE $\gamma$-ray regime. In 2006, besides the central $\gamma$-ray emitter (HESS J1745-290~\cite{Aharonian:2004wa,Aharonian:2009zk}) and the supernova/pulsar wind nebula G0.9+0.1~\cite{Aharonian:2005br}, the \HESS\ experiment discovered a diffuse $\gamma$-ray emission at energies of ($0.2-20$) TeV. This emission spreads through  the central 200 pc along the Galactic plane, with a spatial extension of about $\pm$0.3$^{\circ}$ in Galactic latitude, likely to be correlated with the Central Molecular Zone~\cite{Aharonian:2006au}. In order to avoid contamination from the strong astrophysical background from the above-mentioned VHE $\gamma$-ray emissions, Galactic latitudes with $|b|<0.3^{\circ}$ are excluded in the RoIs considered in this study  as briefly stated in Sec.~\ref{RoIsdef}.

On the other hand, at lower energies (below roughly 500 GeV), the \FERMI\ satellite has measured a diffuse $\gamma$-ray emission~\cite{FermiLAT:2012aa,Ackermann:2014usa}, originated mostly by the $\pi^0$ decays from  proton-proton collisions in the interstellar medium. In our RoIs, the GDE is well described by a power-law spectrum extending up to roughly 500 GeV and an accurate mapping of this emission in the inner $5^\circ$ can be in principle inferred from Ref.~\cite{fermibck}. Nevertheless, since above the \CTA\ threshold the $\gamma$-ray contamination is not fully understood, we will limit ourselves by considering two extreme choices of the GDE background. In one  case, we do not consider at all the GDE in the \CTA\ energy window.  In another case  we will assume an {\it isotropic} flux all over the RoIs using an averaged normalization for the GDE spectrum. In particular, we take  the GDE flux per steradian   $\rm d\Phi^{\rm GDE}_{\gamma}/({\rm d}E_{\gamma}\rm d\Omega)$ coming from the public P7V6 GDE model within one degree from the GC by the \FERMI\ collaboration and we extrapolate it to higher energies to cover all the accessible energy range of \CTA. This latter choice is conservative because we are taking the higher $\gamma$-ray contamination in all RoIs and therefore the \CTA\ sensitivity to  DM signals will be maximally deteriorated. Indeed, it is reasonable to expect that: ($i$) the GDE we are using is bigger than the expected one in the first RoI (its normalization is in fact taken from a region with the same angular aperture of the first RoI, without however eliminating the rectangular part which cuts the Galactic plane); ($ii$) the GDE decreases in the outer RoIs (see e.g.~Ref.~\cite{Silverwood:2014yza} where, in their {\em optimistic scenario},  an exact knowledge is assumed of the spatial morphology of the GDE); ($iii$) the GDE is completely absorbed at high energy due to  photon-photon scattering and pair production on ambient photon background radiation (e.g.~the \FERMI\ collaboration has recently reported a possible energy cut-off around 800 GeV, in the spectrum of the isotropic diffuse emission~\cite{Ackermann:2014usa}). 

Having at our disposal the  flux, the number of GDE background events is again given by 
\beq
N_{\gamma, ij}^{\rm GDE}= T_{\rm obs} \int_{\Delta E_j}  \hspace{-.15cm}  \ud E_\gamma \, \frac{\ud \Gamma_{\gamma, i}^{\rm GDE}}{\ud E_\gamma} \ , \qquad \qquad  \frac{\ud \Gamma_{\gamma, i}^{\rm GDE}}{\ud E_\gamma}= \int_{-\infty}^{+\infty} \hspace{-.15cm} \ud E_\gamma' \, \frac{\ud \Phi_{\gamma}^{\rm GDE}}{ \ud E_{\gamma}\ud \Omega}(E_\gamma') \Delta\Omega_i \, \mathcal A^{\gamma}_{\rm eff}(E_\gamma') \, \frac{e^{-\frac{(E_\gamma-E_\gamma')^2}{2\sigma^2(E_\gamma')}}}{\sqrt{2\pi\sigma^2(E_\gamma')}}  \ ,
\eeq
where $\ud \Gamma_{\gamma, i}^{\rm GDE}/\ud E_\gamma$ is the differential count rate. The blue dotted lines in the left and central panels of Fig.~\ref{fig:plot_S_B} show the spectral morphology of $E_\gamma^2 \cdot \ud \Phi_{\gamma,1}^{\rm GDE}/\ud E_\gamma$ and  $\ud \Gamma_{\gamma, 1}^{\rm GDE}/\ud E_\gamma$, respectively. From the central panel, we clearly see that the GDE will increases the total background up to a few TeV energies. This will impact our expected sensitivity especially for low DM masses as we will show in Sec.~\ref{sec:results}. 


The filled-blue stars in the right panel of Fig.~\ref{fig:plot_S_B}, shows instead the spatial morphology of the total GDE rate summed over all energy bins above 30 GeV ($\Gamma_{\gamma, i}^{\rm GDE}=1/T_{\rm obs} \sum_j N_{\gamma, ij}^{\rm GDE}$)). Since, we assume that the GDE is isotropic, the spatial dependence of this background is clearly contained in $\Delta\Omega_i$.  

\section{Analysis Methodology}
\label{sec:obs}

\begin{figure}[t]
\centering
\includegraphics[width=.45\textwidth]{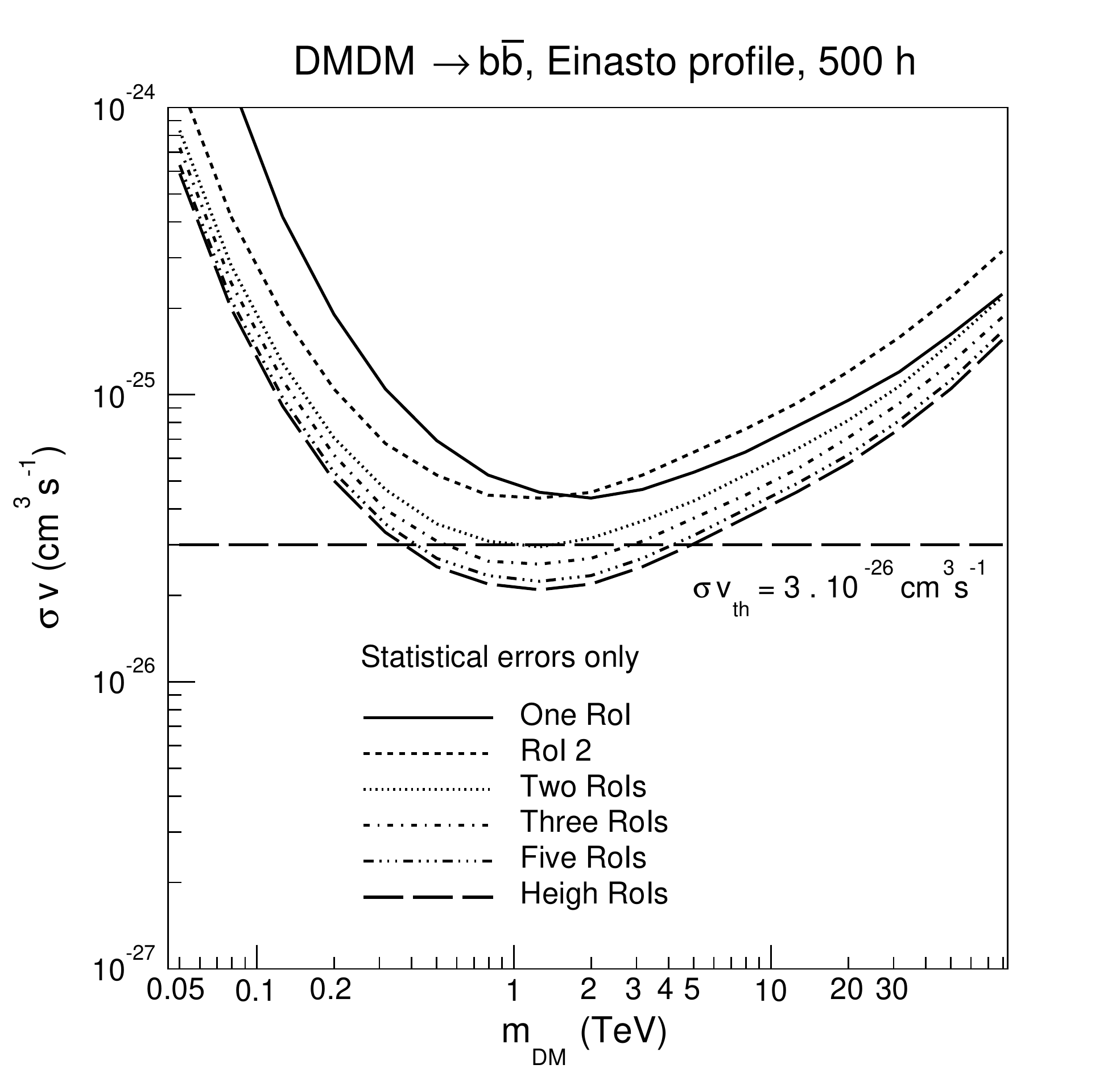} 
\includegraphics[width=.45\textwidth]{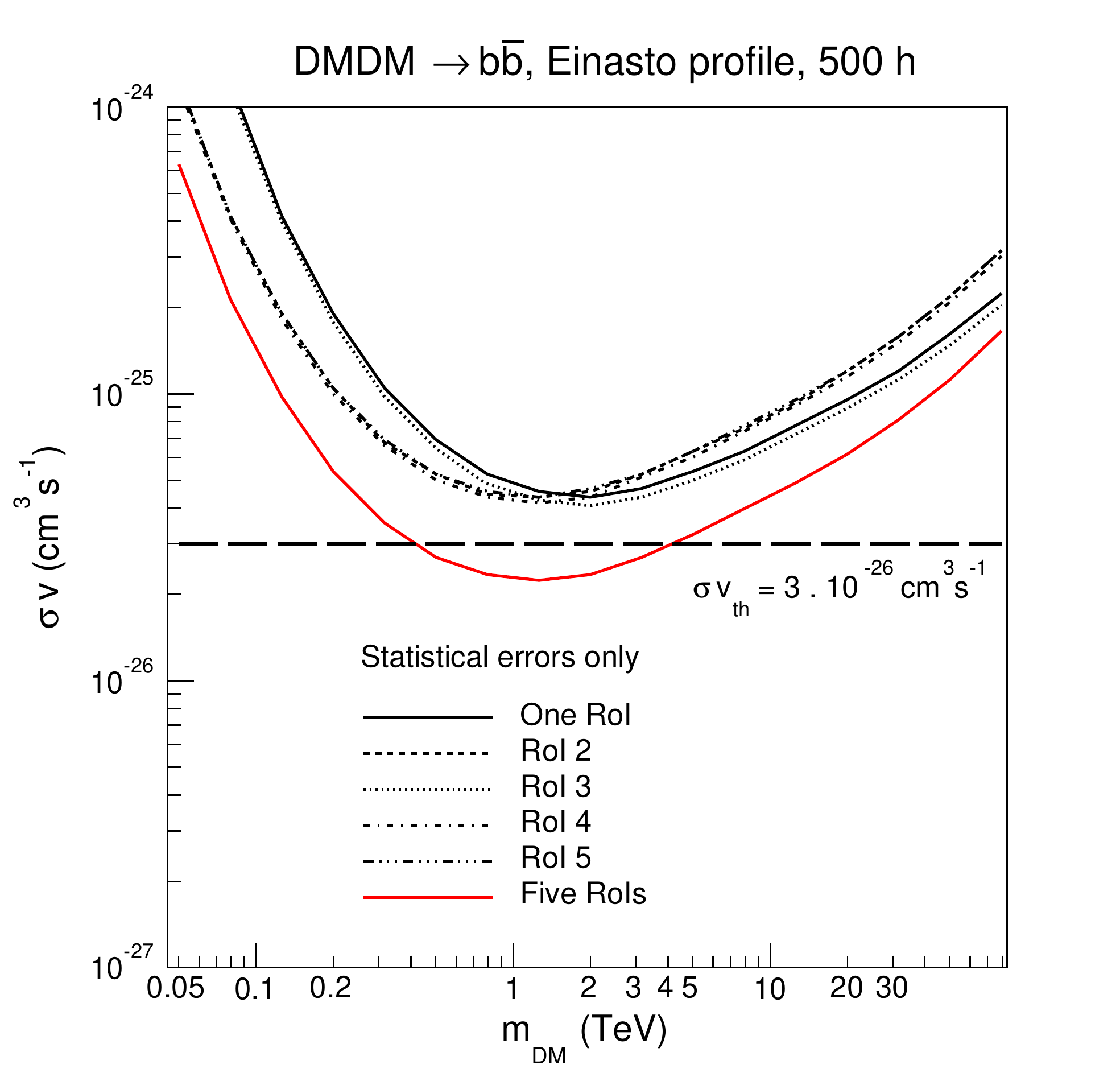} 
\includegraphics[width=.45\textwidth]{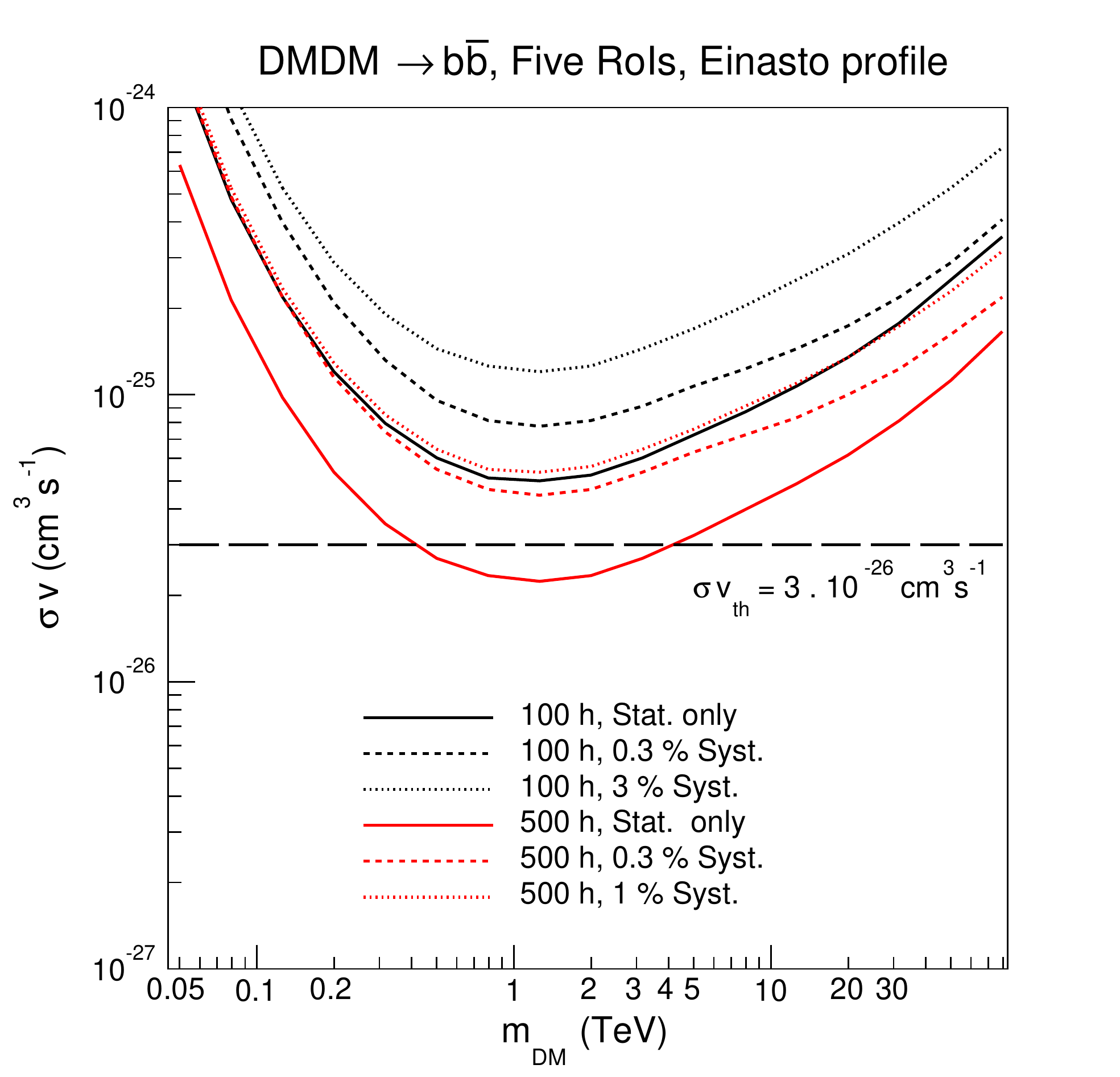}
\includegraphics[width=.45\textwidth]{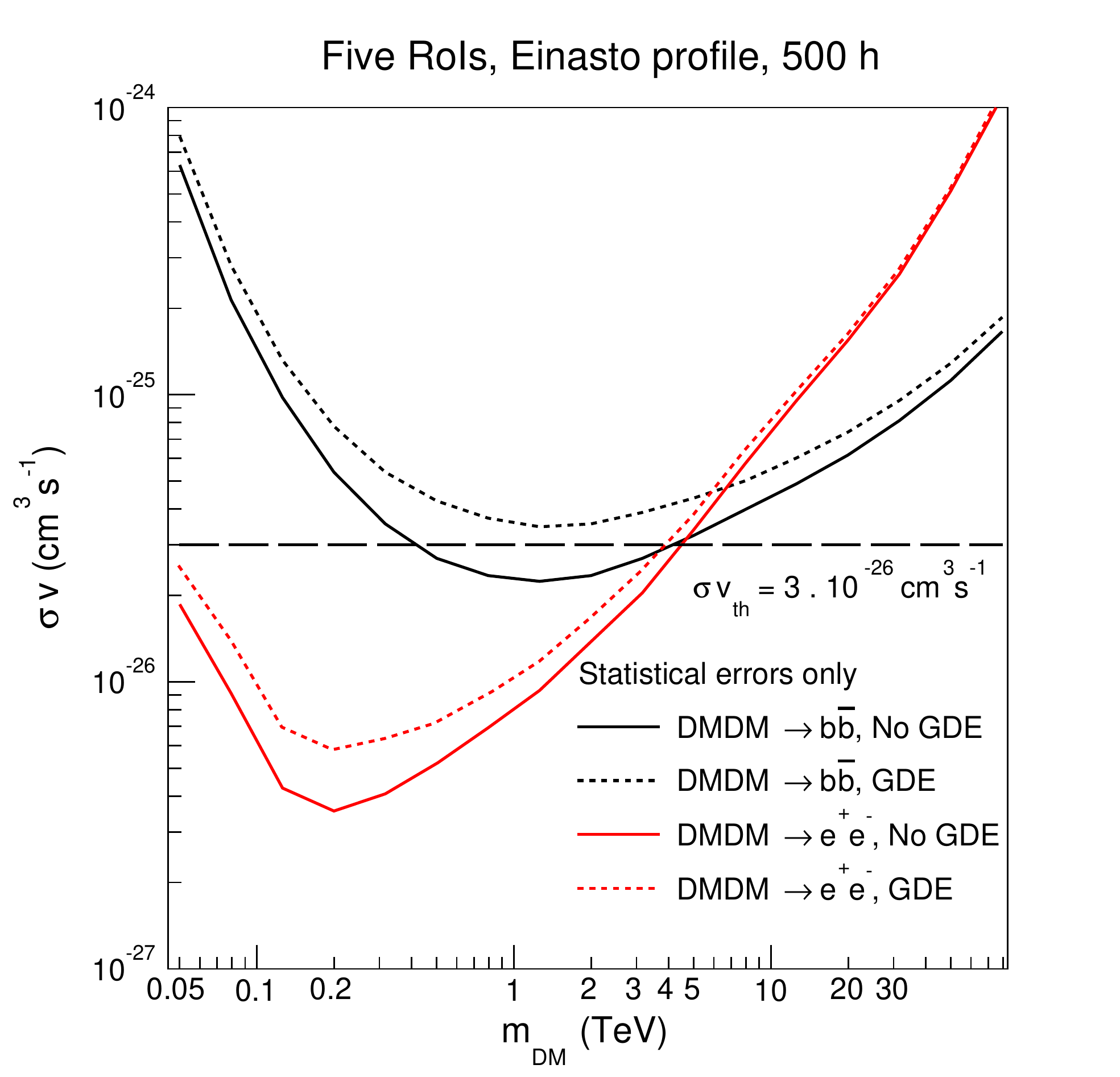} 

\caption{\small \CTA\ sensitivity to DM annihilation  in the ($\mDM, \sigmav$) plane.  Unless otherwise stated, our morphological analysis employs five adjacent RoIs, the exposure in each RoIs is 500 h, the energy threshold is 30 GeV and only the statistical uncertainties are taken into account. 
{\em Upper-left panel:} Improvement of the \CTA\ sensitivity for the DM DM $\rightarrow \bar b b$ channel due to our multi-bin morphological method (two, three, five and eight RoIs) compared to the case where only the energy spectral information either in the first (solid black line) or in the second (dashed black line) RoI is taken into account. 
{\em Upper-right panel:} \CTA\ sensitivity for the DM DM $\rightarrow \bar b b$ channel when only the spectral information in the individual RoIs (RoI 1 to RoI 5) is considered. For a sake of comparison, we also show how our multi-bin morphological method (solid red line) ameliorates the \CTA\ sensitivity.
{\em Bottom-left panel:} Degradation of the \CTA\ sensitivity for annihilating DM into $\bar b b$ pairs due to possible systematic errors in the rich observational datasets. Observation times of 100 h (black lines) and 500 h (red lines) assuming different values of systematics: 0.3\% (dashed lines), 3\% (dotted lines), only statistical fluctuations (solid lines) are taken into account.
{\em Bottom-right panel:} Impact of an ``extreme'' GDE added on top of the CR background, on the \CTA\ sensitivity. The sensitivity is shown for annihilating DM into $\bar b b$ (back lines) and $e^+e^-$ (red lines) pairs with (dotted lines) and without  (solid lines) the irreducible GDE background.
See the text for further details.}
\label{fig:plot_sigmav}
\end{figure}

As stated in Sec.~\ref{sec:cta}, observations by IACTs in high energy $\gamma$-ray astronomy usually employ two regions  on the sky: one in the direction where the signal is expected to be (ON region), and one for the background determination (OFF region).  As follows, we will implement  the Li\&Ma notations~\cite{Li:1983fv}  used for the above mentioned ON-OFF method, by carrying out a full likelihood analysis which uses the  expected spectral and spatial morphology of the DM signals.

\subsection{Sensitivity calculation methodology}

The statistical method to derive the sensitivity is based on a likelihood ratio statistical test. In order to take full advantage of the specific spatial and spectral  features in the DM signal (see Figs.~(\ref{fig:En2_exp},\ref{fig:plot_S_B})), the total likelihood for a given $\mDM$ is obtained from the product over the spatial bins $i$ and the energy bins $j$ of the individual Poisson likelihoods. It explicitly writes
\begin{equation}
\mathcal{L}  (\mDM,\langle \sigma v \rangle) = \prod_{i,j} \mathcal{L}_{ij}  (\mDM,\langle \sigma v \rangle) \ ,
\end{equation}
where the spectral part of the likelihood runs from the energy threshold of 30 GeV up to $\mDM$, while  the spatial part runs over the five  RoIs described in Sec.~\ref{RoIsdef}. Following Ref.~\cite{2011EPJC711554C}, the individual likelihood  is then given by
\begin{equation}\label{Likelihood}
{\mathcal{L}_{ ij}(N_{\gamma}^{\rm S},N_{\gamma}^{\rm B}|N_{\rm ON},N_{\rm OFF}) = \frac{(N_{\gamma, ij}^{\rm S}+N_{\gamma, ij}^{\rm B})^{N_{{\rm ON}, ij}}}{N_{{\rm ON}, ij}!}e^{-(N_{\gamma, ij}^{\rm S}+ N_{\gamma, ij}^{\rm B})} \frac{\left(N_{\gamma, ij}^{\rm B}/ \alpha_i \right)^{N_{{\rm OFF},ij}}}{N_{{\rm OFF},ij}!} e^{-N_{\gamma, ij}^{\rm B}/\alpha_i}} \ ,
\end{equation}
where $N_{\gamma, ij}^{\rm S}+N_{\gamma, ij}^{\rm B}$ is the predicted number of events in the $i$-esime RoI and $j$-esime energy bin which corresponds to the observed counts in the ON region $N_{{\rm ON},ij}$. Here $N_{\gamma, ij}^{\rm B}=N_{\gamma, ij}^{\rm CR}+N_{\gamma, ij}^{\rm GDE}$ is the predicted number of background events which is estimated from an OFF region in order to avoid modelling of the background contamination in the considered RoI. 
The observed number of background events in the OFF region is denoted by $N_{{\rm OFF},ij} $ and the parameter $\alpha_i=\Delta\Omega_i/\Delta\Omega_{\rm OFF}$ refers to the ratio between the angular size of the $i$-esime ON region and the OFF one.  From Eq.~\eqref{Likelihood}, one can then infer that $N_{{\rm ON}, ij}$ and $N_{{\rm OFF}, ij}$ correspond to Poisson realizations with mean $N_{\gamma, ij}^{\rm S}+N_{\gamma, ij}^{\rm B}$ and  $N_{\gamma, ij}^{\rm B}/\alpha_{i}$ respectively. The background is estimated beyond the fifth RoI in a region five times larger. In this case the ratio between the fifth RoI and OFF region is $\alpha_i$ = 0.2. We assume a conservative constant value of $\alpha$ even if the value of it is smaller for the inner four RoIs. For our statistical analysis we will adopt the  likelihood ratio  test statistic  ${\rm TS}=-2 \ln(\mathcal{L}(\mDM,\langle \sigma v \rangle)/\mathcal{L}_{\rm max}(\mDM,\langle \sigma v \rangle))$ which follows an approximate $\chi^2$ distribution with one degree of freedom. Values of TS higher than 2.71 are excluded at a 95\% Confidence Level (C.L.). 

\smallskip
Another possible approach, which has been used e.g.~in~Ref.~\cite{Silverwood:2014yza}, is to consider only ON regions in the likelihood. In this case, the Poissonian term for $N_{{\rm OFF},ij}$ in the right-hand side of Eq.~\eqref{Likelihood}, is not considered at all. Therefore the background determination relies on a careful modelling in the RoIs. Although this approach allows to put stronger constraints, it crucially depends on the residual background modelling accuracy. In our multi-RoI approach,  the background is computed from data taken in real observation conditions and therefore allows for more accurate background determination.

\subsection{Impact of background systematics}
\CTA\ observations towards the inner Galactic halo will provide statistically rich observational datasets. This datasets result  from the combination of multiple observations with distinct observational parameters that may introduce specific observational systematics. 

Since the systematic errors may be the limiting factor in the overall error budget for CTA, an assessment of the impact of them in the CTA sensitivity is in order.  A possible way to investigate the impact of such uncertainties  is to introduce in the likelihood  a Gaussian nuisance parameters~\cite{2011EPJC711554C} such as the individual likelihood writes
\begin{equation}
{\mathcal{L}_{ ij}(N_{\gamma}^{\rm S},N_{\gamma}^{\rm B}, \beta|N_{\rm ON},N_{\rm OFF}) = \frac{e^{-{\frac{(1-\beta_{ij})^2}{2\sigma^2_{ij}}}}}{\sqrt{2\pi}\sigma_{ij}} \frac{\beta_{ij}^{N_{{\rm ON},ij}}(N_{\gamma, ij}^{\rm S}+N_{\gamma, ij}^{\rm B})^{N_{{\rm ON},ij}}}{N_{{\rm ON},ij}!} e^{-\beta_{ij}(N_{\gamma, ij}^{\rm S}+N_{\gamma, ij}^{\rm B})}  \frac{\left(N_{\gamma, ij}^{\rm B}/ \alpha_i \right)^{N_{{\rm OFF},ij}}}{N_{{\rm OFF},ij}!} e^{-N_{\gamma, ij}^{\rm B}/\alpha_i}} \ ,
\end{equation}
where $\beta_{ij}$ acts as a normalization parameter and $\sigma_{ij}$ is the width of the Gaussian function. An accurate determination of the spatial and energy dependencies of the systematic level is beyond the scope of this study, and we will keep $\sigma_{ij}$ fixed for all spatial and energy bins. We can determine the maximum likelihood value of it  by solving $\ud \mathcal{L}/\ud \beta_{ij} = 0$ in the $i$-esime RoI and $j$-esime energy bin, for a given set of $N_{\gamma,ij}^{\rm S}$, $N_{\gamma,ij}^{\rm B}$, $N_{{\rm ON},ij}$ and $N_{{\rm OFF},ij}$. 
In order to evaluate the impact of systematics in the CTA sensitivity, we will then consider several plausible values of the Gaussian width  $\sigma$ for IACT observational data. In particular we will vary it from 0.3\%  to 3\% in order to compare with the case where only the statistical uncertainty is taken into account.

\section{Results}
\label{sec:results}
We now show our results in terms of 95\% C.L. sensitivity limits on DM annihilation, in the usual ($\mDM, \sigma v$) plane. 
We focus on several particle-antiparticle annihilation modes (DM DM $\rightarrow e^+e^-,  \mu^+\mu^-, \tau^+\tau^-, b\bar{b}, t\bar{t}$ and $W^+W^-$) and in a broad range of DM masses (from 30 GeV up to 80 TeV). For all the  channels, we include the ICSs of energetic $e^\pm$ produced by annihilating DM on the ambient photon background which  is particularly relevant for  $e^+e^-$ and $\mu^+ \mu^-$ modes (see the red lines in Fig.~\ref{fig:En2_exp}).  We summarize our main results in Fig.~\ref{fig:plot_sigmav}. In particular we find that:

\begin{itemize}
\item[$\diamond$] {\em The ICS emission substantially increases the CTA sensitivity for the leptonic channels.}  In Fig.~\ref{fig:En2_exp}, we show the spectral features of the ICS emissions. As it is apparent, the total fluxes receive a substantial contribution for photon energy just below the DM mass in case of  leptonic channels (especially for the DM DM $\rightarrow e^+e^-$ and $\mu^+\mu^-$ modes). As a consequence, since the IC secondary emission is  well inside the \CTA\ energy window, the sensitivity to those channels is largely ameliorated. 

\item[$\diamond$] {\em A morphological analysis ameliorates the CTA sensitivity.} Assuming again an observation time of 500 h, in the upper-left panel of Fig.~\ref{fig:plot_sigmav}  we show  the improvement of the sensitivity for the $\bar b b$ channel, by using the multi-bin morphological method compared to the case where only the energy spectral information {either in  the first RoI (solid black line) or in the second one (dashed black line)} is used. The constraints inferred with this approach are more stringent by a factor of a  few with respect  to those obtained considering only the spectral information in any single RoI. Indeed, from the upper-right panel of the same figure, one can quantify the impact of our morphological analysis by comparing the different black lines (one RoI, RoI 2, 3, 4, 5 alone) with the red one (combination of all the relevant RoIs considered in our analysis). As one can see all the constraints coming from individual region, are a factor $\mathcal O(2)$ less stringent than the one inferred with a multi-bin morphological method. As shown in the right panel of Fig.~\ref{fig:plot_S_B}, this is mainly due to the fact that the spatial dependence of the DM signal flatten above roughly the third RoI, while the CR background contamination increases  with the size of the $i$-esime region. 

\item[$\diamond$] {\em The systematic uncertainties  deteriorate the CTA sensitivity.} In the bottom-left panel of Fig.~\ref{fig:plot_sigmav}, we show the impact on the \CTA\ sensitivity for annihilating DM into $b\bar{b}$ pairs due to possible systematic errors in the rich observational datasets.  We use our morphological analysis (five RoIs) with observation times of 100 h (black lines) and 500 h (red lines) assuming different values of systematics: 0.3\% (dashed lines), 1\% (red dotted line), 3\% (black dotted line), only statistical fluctuations (solid lines). We can see that  the sensitivity is deteriorated over all the mass range for a given observational time. In particular,  introducing a systematic errors of 0.3\%(3\%) for 100 hours, deteriorates the sensitivity of a factor 1.5 (see Fig.~\ref{fig:plot_sigmav}). For multi-TeV DM masses, the impact is reduced because at higher energy the $\gamma$-ray datasets are dominated by the statistical errors. We stress that the systematics must be controlled at the level of  0.3\% or better to allow for a substantial sensitivity improvement with a 500 h observation time.

\item[$\diamond$] {\em Our extreme choice of the  Galactic Diffuse Emission degrades the CTA sensitivity}. Assuming an observation time of 500 h, in the bottom-right panel of Fig.~\ref{fig:plot_sigmav} we assess the \CTA\ sensitivity for annihilating DM into $\bar b b$ (back lines) and $e^+e^-$ (red lines) pairs once an ``extreme'' GDE is added to the CR background. As commented upon in Sec.~\ref{sec:GDE}, we consider an {\em isotropic} GDE coming from the inner Galactic halo. We can see that the \CTA\ sensitivity is significantly deteriorated below DM masses of few TeV (dashed lines) with respect to the scenario where the GDE is not considered at all (solid lines). On a more specific level, even with our ``extreme''  choice of the GDE, the \CTA\ sensitivity still probe cross section below the thermal value for the  $e^+e^-$ channel. For the hadronic ones (e.g.~$\bar b b$ mode), the \CTA\ sensitivity is degraded of a factor 2 making the reach of the thermal cross section no longer possible. Nevertheless, since we  assume that the GDE is {\em isotropic}, it is worth stressing once again that we are overestimating the $\gamma$-ray contamination in all RoIs. In fact, if we consider an accurate mapping of the GDE in our RoIs (like the one used in Ref.~\cite{Silverwood:2014yza} in their {\em optimistic scenario}), we find that the impact of the GDE in the final results is not very pronounced.  This is due to the fact that in all the regions used in our analysis, the diffuse $\gamma$-rays contamination is smaller than the residual CR background extracted from a full \CTA\ Monte Carlo simulation.

\end{itemize}

\begin{figure}[t]
\centering
\includegraphics[width=.49\textwidth]{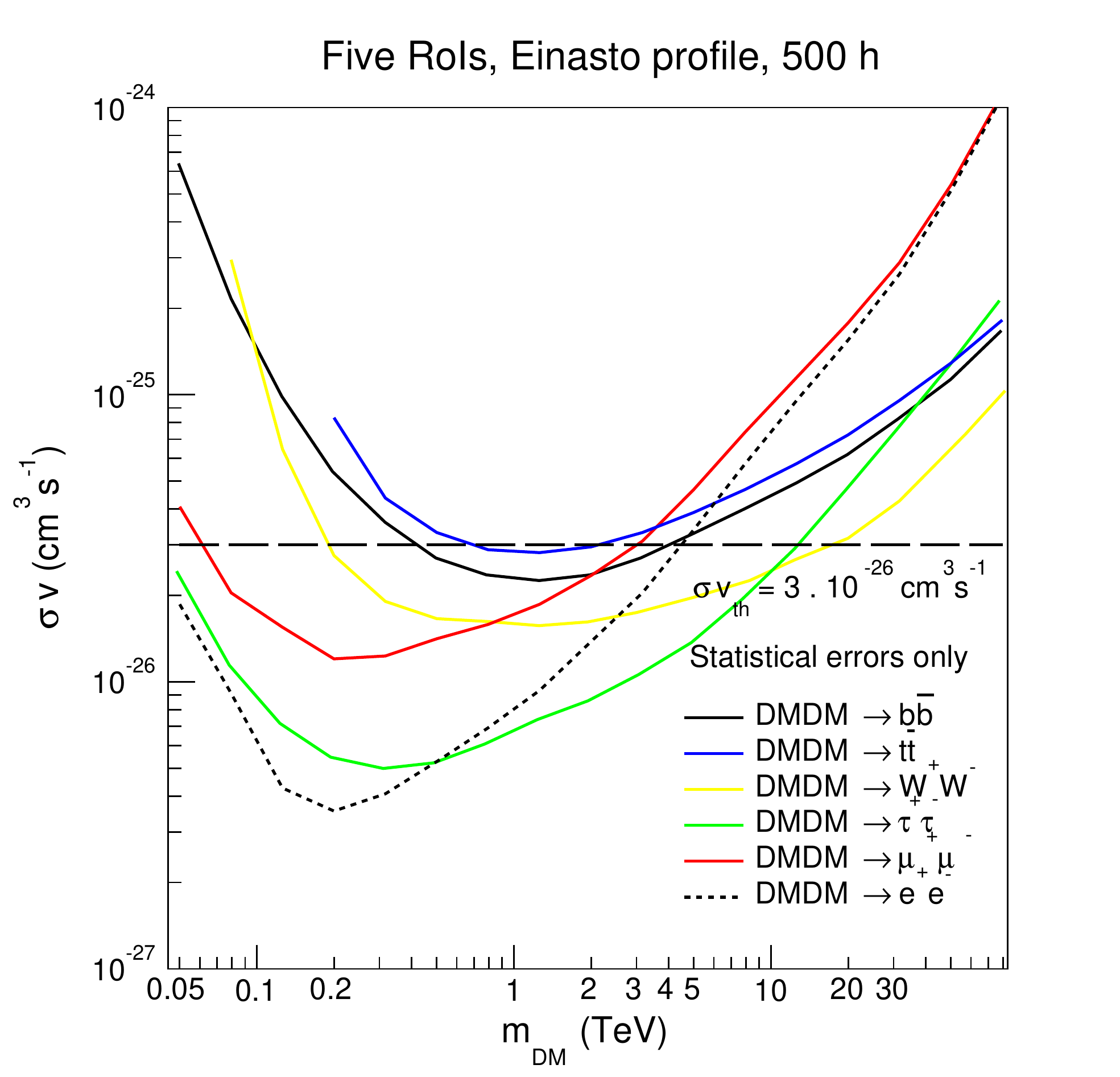} \
\includegraphics[width=.49\textwidth]{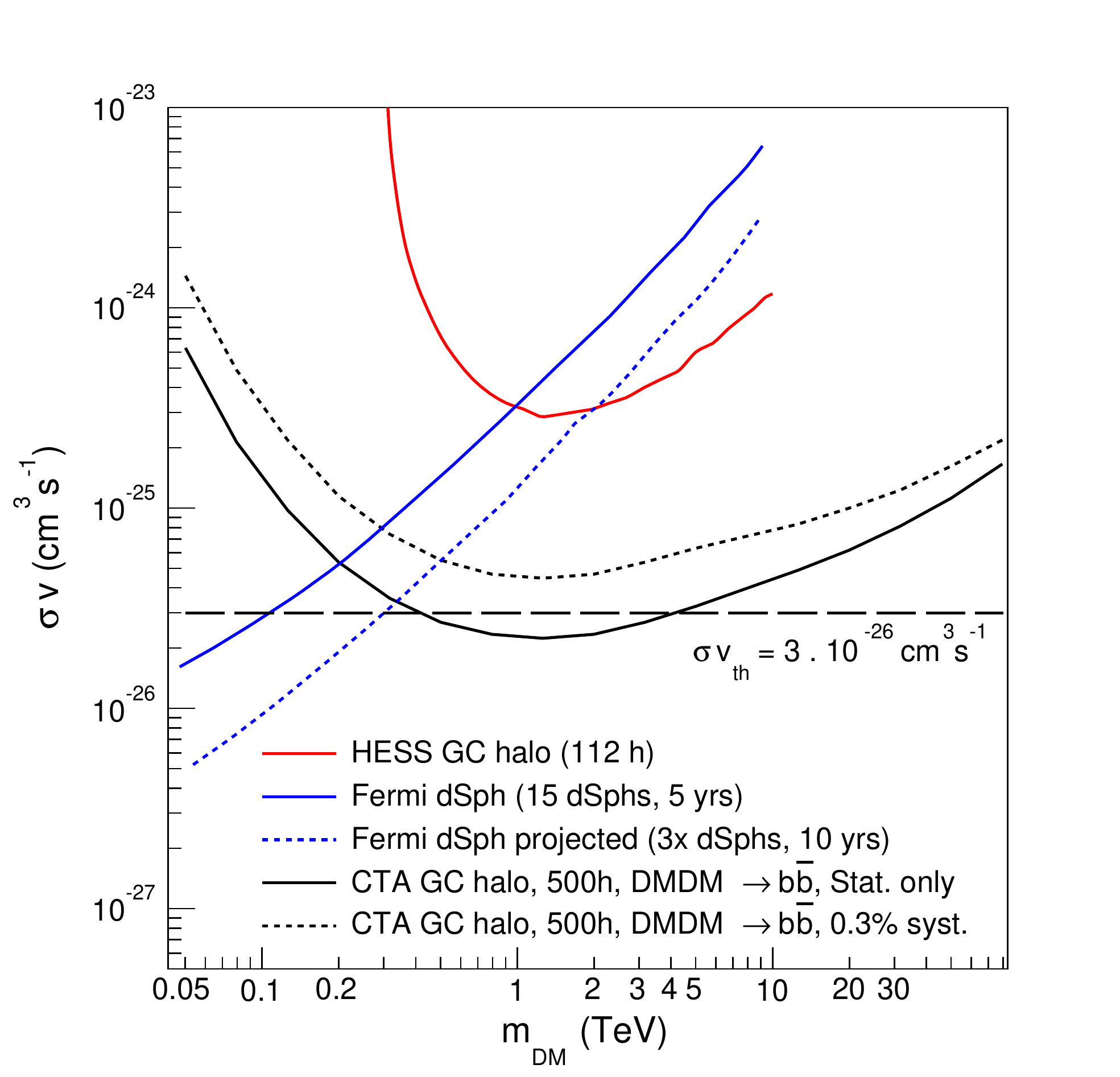}
\caption{\small Unless otherwise stated, the {\em most optimist} limits in the ($\mDM, \sigmav$) plane in which neither the systematic uncertainties in the datasets nor the GDE are taken into account. {\em Left panel:} \CTA\ sensitivity to DM annihilation for various primary channels (DM DM $\rightarrow b\bar{b}$ in black, $\bar t t$ in blue, $W^+W^-$ in yellow, $\tau^+\tau^-$ in green, $\mu^+\mu^-$ in red and $e^+e^-$ in dashed black) together with the reference value of the thermal cross section (long-dashed black line). The sensitivity is computed for a 500 h flat exposure over the five RoIs used in our analysis assuming a 100\% branching ratio in each annihilation channel. 
 {\em Right panel:}  \CTA\ sensitivity to DM annihilation for the DM DM $\rightarrow b\bar{b}$ channel (black line) compared to the most constraining limits to date. In particular, we report the \HESS\ limit for 112h of observations of the GC region (solid red line) and the \FERMI\ limit coming from 5 years of  observations of 15 dSphs (solid blue line). An estimate of the projected sensitivity of \FERMI\ for 45 dSphs and 10 years of observation time is also shown (dotted blue line). For a sake of comparison, we also report the \CTA\ sensitivity assuming a reasonable 0.3\% of systematics (dashed black line). See the text for further details.}
\label{fig:plot_channels}
\end{figure}

In Fig.~\ref{fig:plot_channels}, unless otherwise stated, we show the {\em most optimistic} limits in the ($\mDM, \sigmav$) plane in which neither the systematic uncertainties in the datasets nor the GDE are taken into account.

In particular, in the left panel of Fig.~\ref{fig:plot_channels} we show the \CTA\ sensitivity to DM annihilation assuming different  channels (DM DM $\rightarrow e^+e^-,  \mu^+\mu^-, \tau^+\tau^-, b\bar{b}, t\bar{t}$ and $W^+W^-$), and an observation time of 500 h. Focussing first on the purely leptonic  channels, we find that \CTA\ would be able to exclude annihilation cross-sections well below the thermal value. On a more specific level, the best sensitivity is obtained for the DM DM $\rightarrow e^+e^-$ mode ($\sigmav \lesssim 5\times 10^{-27}$ cm$^3$/s for $\mDM\simeq 200$ GeV) rather than $\mu^+\mu^-$ and $\tau^+\tau^-$, since the original $e^\pm$ population is produced at higher energies, and therefore the secondary ICS emission is well inside the \CTA\ energy window (see the upper raw in Fig.~\ref{fig:En2_exp}). For the hadronic and $W^+W^-$ channels, we get the same qualitative feature of the exclusion limits modulo a factor of $\mathcal O$(few) in the normalization. This can be explained from the fact the $\gamma$-ray spectrum arising from the fragmentation of sufficiently heavy hadronic SM particles is quasi-universal. In particular, for the DM DM $\rightarrow b\bar{b}$ channel the thermal value of the cross section can be probed in the TeV mass range, where the best sensitivity is achieved at $\sigmav \simeq 2 \times 10^{-26}$ cm$^3$/s for $\mDM\simeq 1$ TeV.

\medskip 
We comment here on the relative strength of our constraints  for the DM DM $\rightarrow \bar b b$ channel in Figs.~(\ref{fig:plot_sigmav},\ref{fig:plot_channels}), with respect to the best limits to date obtained from either other analyses or  targets.
For a sake of comparison, in the right panel of Fig.~\ref{fig:plot_channels}, we also report the \CTA\ sensitivity assuming a reasonable 0.3\% of systematics (dashed black line). 

With respect to the \HESS\ limits from  112 h of observations in the central region of the Galactic halo~\cite{Abramowski:2011hc} (red line in the right panel of Fig.~\ref{fig:plot_channels}), the bounds derived here allows to probe a larger range of DM masses (from 50 GeV up to 80 TeV). In particular the \CTA\ sensitivity with no GDE and systematics will be roughly a factor 10 more sensitive for DM masses around 1 TeV. 

With respect to the most recent stacking analysis on 15 dwarf Spheroidal galaxy (dSph) observations from \FERMI~\cite{Anderson14} (solid blue line in the right panel of Fig.~\ref{fig:plot_channels}), \CTA\ becomes competitive for energies above 100 GeV and it overtakes the \FERMI\ constraint for DM masses above 200 GeV.  Furthermore, it is important to point out that \CTA\ will still provide stronger limits above roughly 500 GeV, if  the very optimistic scenario of 45 observed dSphs for 10 years \FERMI\ observations~\cite{Anderson14}  will be considered (dashed blue line in the right panel of Fig.~\ref{fig:plot_channels}). Hence, together with the optimistic \FERMI\ observations of dSphs, \CTA\ will be able to survey thermal DM candidates in a broad range of masses (from few tens of GeV up to several tens of TeV). 

\smallskip
With respect to previous \CTA\ projected limits from Refs.~\cite{Doro:2012xx, Wood:2013taa, Pierre:2014tra},  our results  are more conservative. In particular, in contrast with these works, we include both the systematic uncertainties in the datasets and an ``extreme''  GDE on top of the CR background. We  find that the shape of the bounds as a function of the DM mass is very different. This is mainly due to the fact that we use both spectral and spatial information in the likelihood. The impact  of including spectral information was pointed out  in Ref.~\cite{Pierre:2014tra}, whose limits have in fact a similar shape compare to our bounds.   The fact that our limits with no GDE and systematics are instead less stringent than those obtained in Ref.~\cite{Doro:2012xx}, is mainly due to the fact that we use a less steep profile rather than a profile inferred from the Aquarius simulation which include a significant boost due to substructure. Indeed, in their ON region, which almost coincide with our first RoI, the value of the $J$-factor based on the Aquarius simulation is $4.68\times 10^{22}$ GeV$^2$/cm$^5$, while in our case is around $1.42\times 10^{21}$ GeV$^2$/cm$^5$.  This yields a difference in the flux compare to our work of a factor $\sim 33$. Nevertheless, since we  use a morphological analysis which improves the \CTA\	 sensitivity of roughly a factor 2,  we should multiply the limits in Ref.~\cite{Doro:2012xx} by 33/2  for a sake of comparison. 

\smallskip
With respect to Ref.~\cite{Silverwood:2014yza}, which is similar to our work in many aspects, we find again that our results are more conservative. This is mainly due to the fact that the parameters of the Einasto profile used in our analysis are different. Considering their  $J$-factor (in Ref.~\cite{Silverwood:2014yza} the $J$-factor in our first RoI\footnote{Notice that in Refs.~\cite{Silverwood:2014yza} the value of the $J$-factor is quoted in a region slightly bigger than ours. In particular, they considered a circle centered in the GC with an aperture of $1.3^\circ$, minus a rectangular region which cuts the galactic plane.} (see Tab.~\ref{tableJ}) is around $4.61\times 10^{21}$ GeV$^2$/cm$^5$), and the fact that we implement, in a different way, a morphological analysis  in the likelihood (improvement of the CTA sensitivity of a factor 2), we find however that the bound for the DM DM$\rightarrow b\bar{b}$ channel for 100 h observational time is in well agreement with our own. 
The only notable difference has been found for DM masses above roughly 1 TeV, since, as pointed out many times, we use the most up-to-date CR background obtained from a full \CTA\ Monte Carlo simulation. This gives a larger contamination above roughly 1 TeV compare to the analytic estimation of the CR background used in Ref.~\cite{Silverwood:2014yza}. As a consequence, our bounds are less stringent. The fact that our background is bigger at larger energies is also relevant for the discussion related to the GDE. In fact we find that only if we choose an ``extreme'' GDE, the constraints are significantly affected. If we use the accurate mapping of the GDE used in Ref.~\cite{Silverwood:2014yza} in their {\em optimistic scenario}, we only report a barely modification of the \CTA\ sensitivity. Finally, concerning the impact of the systematics errors on the constraints, we find an almost perfect agreement with that presented in Ref.~\cite{Silverwood:2014yza}.

\section{Summary}
\label{sec:sum}

We discuss the future \CTA\ sensitivity to DM annihilations in several channels and over a range of DM masses from 50 GeV up to 80 TeV. For all channels we include the ICS emissions which yield a substantial contribution to the overall $\gamma$-ray flux, especially for the leptonic channels.   We improved the analysis over previous work by: $i)$ implementing a spectral and morphological analysis in the $\gamma$-rays emission; $ii)$ taking into account the most up-to-date CR background obtained from a full \CTA\ Monte Carlo  simulation and a description of the GDE on top of this; and $iii)$ including the systematic errors in the rich observational \CTA\ datasets. 

\smallskip
We showed that our morphological analysis with five RoIs substantially  improves the \CTA\ sensitivity by roughly a factor of 2. In particular, for the hadronic channels, we found that the \CTA\ with an uniform exposure of 500 h of observations, will be able to probe thermal values of the annihilation cross-section over a broad range of DM masses, if the systematics uncertainties in the datasets will be controlled better than the percent level. For the leptonic channels, and in particular for the DM DM $\rightarrow e^+e^-$ mode, we found  that the bounds are instead well below the thermal value of the annihilation cross section. In this case, even with larger systematics, thermal DM candidates up to masses of few TeV will be easily probed. 

\smallskip
In contrast  with Ref.~\cite{Silverwood:2014yza}, we found that the inclusion of the GDE in the overall background produces a sizeable effect on the \CTA\ sensitivity, only if one assumes an {\em isotropic} $\gamma$-rays contamination extracted from the inner galactic halo. With a more physical choice, which includes  a proper spatial dependence of the GDE, we only found a minimal impact in the \CTA\ sensitivity. This is due to the fact that our CR background is larger than the one analytically inferred in Ref.~\cite{Silverwood:2014yza}. Therefore, apart from an  ``extreme'' choice of the GDE,  the degradation of the \CTA\ sensitivity, especially at high energy (above roughly 1 TeV), is mostly due to the CR background contamination. 

\smallskip
In summary, in order to probe the thermal value of the annihilation cross-section over a broad range of DM masses, deep observations of the GC over several degrees in radius (at least up to 5 degrees) are required with uniform exposure and the best possible control of systematic uncertainties. Under these conditions, \CTA\ will give crucial information for TeV-ish WIMP searches in the next decade. Furthermore, in the {\em optimistic scenario} where the \LHC\ will discover new physics in which thermal DM candidates are present, \CTA\ will be probably the only player that could cross-check such possible results against an astrophysical environment.

\acknowledgements
P.P. and J.S. acknowledge the support of the European Research Council project 267117 hosted by Universit\'e Pierre et Marie Curie-Paris 6, PI J.~Silk.

\clearpage
\bibliography{bibl}

\end{document}